\documentclass[12pt,authoryear]{elsarticle}

\usepackage{amssymb}

\usepackage[english]{babel}
\usepackage[T1]{fontenc}
\usepackage[utf8x]{inputenc}
\usepackage{a4wide}
\usepackage[centertags]{amsmath}
\usepackage{amsfonts}
\usepackage{amssymb, verbatim}
\usepackage{hyperref}
\usepackage{amsthm, tabularx}
\usepackage{newlfont}
\usepackage{amssymb,amsbsy,times,fancyhdr,color}
\usepackage{amsmath}
\usepackage{epstopdf}
\usepackage{multirow}
\usepackage{latexsym}
\usepackage{color, url}
\usepackage[normalem]{ulem}
\usepackage[title]{appendix}
\usepackage[authoryear]{natbib}
\usepackage{setspace}
\usepackage{esint}
\usepackage{float, rotfloat}

\usepackage{eurosym}

\newcommand\obsX{\mathcal{X}}
\newcommand\obsY{\mathcal{Y}}

\begin{document}

\begin{frontmatter}

\title{Bayesian calibration and number of jump components in electricity spot price models}

\author[label1]{Jhonny Gonzalez}
\author[label2]{John Moriarty}
\author[label3]{Jan Palczewski\footnote{Corresponding author. {\em Email address:} {J.Palczewski@leeds.ac.uk}}}

\address[label1]{School of Mathematics, University of Manchester, Manchester M13 9PL, UK}
\address[label2]{School of Mathematical Sciences, Queen Mary University of London,
London E1 4NS}
\address[label3]{School of Mathematics, University of Leeds, Leeds LS2 9JT, UK}

\begin{abstract}
We find empirical evidence that mean-reverting jump processes are not statistically adequate to model  electricity spot price spikes but independent, signed sums of such processes are statistically adequate. Further we demonstrate a change in the composition of these sums after a major economic event. This is achieved by developing a Markov Chain Monte Carlo (MCMC) procedure for Bayesian model calibration and a Bayesian assessment of model adequacy (posterior predictive checking). In particular we determine the number of signed mean-reverting jump components required in the APXUK and EEX markets, in time periods both before and after the recent global financial crises. Statistically, consistent structural changes occur across both markets, with a reduction of the intensity and size, or the disappearance, of positive price spikes in the later period. All code and data are provided to enable replication of results.
\end{abstract}


\end{frontmatter}


\section{Introduction} \label{sec:intro}

Electricity spot markets have multiple fundamental drivers, for example baseload and renewable production \citep{wurzburg2013renewable}. Disturbances in these drivers, such as plant outages and renewable gluts, can clearly have different dynamic characteristics and consequences. 
Since sharp disturbances create spikes in electricity spot prices \citep{seifert_modelling_2007} we may hypothesise that, over time, disturbances in different drivers give rise to spikes with statistically distinguishable directions, frequencies, height distributions and rates of decay. It has recently been demonstrated that electricity spot price formation can evolve over time \citep{brunner2014changes}. Thus we may also hypothesise that the statistical characteristics of electricity price spikes will evolve in step with underlying economic events and developments, such as shifts in demand 
and increasing renewable penetrations.

In this paper we find empirical support for these two hypotheses. To this end we use \emph{multi-factor} electricity spot price models, with multiple superposed mean-reverting components and a seasonal trend \citep{Benth2007}.  This allows  statistical patterns such as mean reversion, seasonality and spikes to be reproduced in modelling. Crucially for the present study, this approach also allows the statistical modelling of multiple spike components with differing frequencies, height distributions, decay rates, and directions (positive or negative). We demonstrate that in some electricity markets two types of positive spike are observed, while other markets require the inclusion of negative spikes. The modelling of negative spikes is an area of emerging interest \citep{fanone2013case} as renewable penetrations, and hence gluts in renewable production, increase.
Finally we document evolution of the statistical spike structure through periods of economic change by comparing two markets across two time periods, one before the recent global financial crises (2000-2007) and another afterwards (2011-2015) and reflect on possible interpretations of the results.

The calibration of multi-factor models 
is a highly challenging task and  existing approaches typically involve making strong {\em a priori} assumptions, such as setting thresholds for jump sizes, which may mask the true statistical structure. Methodologically, we develop a Bayesian approach to calibration based on Markov Chain Monte Carlo (MCMC) methods. This goes beyond previous work by   making minimal assumptions and enables us, for example, to estimate models with multiple spike components acting in the same direction, a feature which is confirmed empirically (in 2001--2006 data from the APXUK electricity spot market). In order to assess the number of mean-reverting jump components required we perform a Bayesian procedure of {\em posterior predictive checking}. 

\subsection{Background and related work}\label{sec:bg}

Econometric models of electricity spot prices have a number of important applications. They provide stochastic models which can be used by traders to analyse financial options on power \citep{Benth2007}, and by power system planners to conduct real options analyses for flexible physical assets such as storage and cogeneration \citep{moriarty2017real, kitapbayev2015stochastic}. Further the pronounced price spikes which characterise spot electricity markets are of central interest to electricity market regulators who monitor and influence the economics of markets, 
aiming for example to prevent perceived abuses of market power \citep{stephenson2001electricity}.

The complexity of electricity spot price models, and multi-factor  models in particular, makes their analysis statistically challenging and has given rise to a substantial literature. A single-factor model including the above stylised features was introduced by \cite{clewlow_energy_2000}. Through the use of a threshold, the single-factor model of \cite{geman_understanding_2006} incorporates two jump regimes: when the price is below the threshold jumps are positive, and when the price exceeds the threshold jumps are negative. 
Beginning with \citet{lucia_electricity_2002} multi-factor  models have expressed the price as a sum of unobservable or {\em latent} processes ({\em factors}) with distinct purposes, for example the modelling of short-term and long-term price variations respectively. Unlike many single factor models, multifactor models do not imply a perfect correlation between changes in spot, future and forward prices, which is consistent with the non-storability of electricity \citep{benth2009information}. The model of \cite{lucia_electricity_2002} has two factors, namely a Gaussian mean-reverting process and an arithmetic Brownian motion (that is, a scaled Brownian motion with drift). Interestingly, while also developing a two-factor model, \cite{seifert_modelling_2007} explicitly refer to the physical origins of various types of jumps. Beyond two-factor models, a simple and flexible multi-factor model with jumps is given in \cite{Benth2007}. Estimation procedures for this model are discussed in \citet{meyer-brandis_multi-factor_2008}, although the latter work adds strong assumptions in order to obtain tractable methods.

The interdependency between parameters in multi-factor  models, in particular, is a challenge to calibration methods. A straightforward approach is to first separate the observed values into factors using signal filtering techniques, in order to subsequently employ classical maximum likelihood estimation. Such methods effectively assume that some of these interdependencies may be neglected, and this approach is taken for example in \citet{meyer-brandis_multi-factor_2008} and \citet{benth_critical_2012}. An alternative is the joint estimation of latent factors, for which there are two leading methodologies in the literature: {\em expectation-maximisation (EM)} and {\em Markov Chain Monte Carlo (MCMC)} methods. While EM produces point estimates for parameters in either a Bayesian or frequentist framework\footnote{Two possible approaches to the calibration of model parameters are commonly referred to as {\em frequentist} and {\em Bayesian}. In the frequentist approach one seeks to derive point estimates of `true' parameter values from the data, for example by finding the maximiser of a likelihood function. An alternative viewpoint is taken in the Bayesian approach, where the unknown parameters are first assigned a probability {\em distribution} representing prior beliefs about their value. This prior distribution is combined with the observed data to produce an updated probability distribution representing the posterior beliefs about the parameters given both the prior and the data.} 
(see, for example, \citet{ryden2008versus}), MCMC is able to generate samples from posterior parameter distributions. Particularly in models with multiple parameters and latent processes, these interdependencies may result in likelihood surfaces and posterior distributions which are rather flat around their maxima. While EM suffers from Monte Carlo errors which amplify the usual difficulties in numerical optimisation for such problems, MCMC estimates the posterior distribution providing an analyst with a more complete picture of the interrelations between parameters.

In related contexts, MCMC has been applied to fit continuous-time stochastic volatility models to financial time series,
where the price is a diffusion process whose volatility is a latent mean reverting jump process or the sum of a number of such processes (called a superposition model). In this line of research a missing data methodology is employed whereby the observed process is augmented with one or more latent marked Poisson processes and the MCMC procedure generates posterior samples in this high dimensional augmented state space. Examples include 
\cite{roberts_bayesian_2004}, \cite{griffin_inference_2006} and \cite{fruhwirth2009bayesian}. 
Since energy prices additionally exhibit jumps directly in their paths,  MCMC has been applied to extensions of these models in which a diffusion process with stochastic volatility is superposed with a jump process, see \citet{2203823} in the context of electricity and \citet{brix2015estimation} for gas prices. Technically the latter two papers estimate a discrete approximation of the models whereas in this study we pursue \emph{exact} inference for continuous time models.

\subsection{Contribution}

From the modelling point of view a novelty of the present study is that the price is a superposition of more than one jump component, each with its own sign, frequency, size distribution and decay rate, along with a diffusion component. This approach acknowledges that the negative price spikes attributable to rapid wind power fluctuations may, for example, have quicker decay than the infrequent larger positive spikes due to major disturbances such as outages of a traditional generation plant. 
The inclusion of multiple jump components also addresses the following problem identified in \citet{2203823} and \citet{brix2015estimation}. In two-factor models jumps of intermediate size must be accounted for either in the diffusion process (forcing unlikely  spikes in the Brownian motion path) or the jump process (implying additional jumps). While the former can lead to an overestimation of volatility in the diffusion process, the latter may result in an overestimation of the intensity of the jump process, which is independent of the jump sizes. The inclusion of a second jump process with its own mean jump size and rate of mean reversion removes this dichotomy, offering an alternative to the inclusion of stochastic volatility in the diffusion process.

Our first methodological contribution is an MCMC algorithm for \emph{exact} Bayesian inference on superposed OU models with diffusion and multiple jump components.  We contrast exact inference with a commonly used estimation procedure using a discrete time model which is an approximation to continuous dynamics. While this approximation is often used for practical reasons including simplified and/or tractable implementation, it is not possible to assess {\em a priori} the extent of the estimation error introduced by the approximation employed. Our MCMC procedure is not based on time discretisation of the model and the inference is therefore exact at the level of distributions. This is in contrast with the work in the aforementioned papers \citep{seifert_modelling_2007, 2203823, brix2015estimation}. Despite the relative simplicity of our multi-factor  model, the MCMC procedure involves a number of challenging issues and in the appendix we provide additional comments and details concerning efficient implementation.

In addition we demonstrate that model {\em adequacy} may also be addressed by our MCMC method. The complexity of electricity spot price models naturally gives rise to parsimony considerations. While multi-factor  models (potentially also including latent volatility processes) offer great flexibility, the potential statistical pitfalls of overly flexible models, for example relating to issues of identifiability and out-of-sample prediction, are well known. In this context, the ability of MCMC to sample whole trajectories from the posterior distribution of the jump processes means in particular that the adequacy of latent variable models may be addressed. We exploit this fact by using the MCMC procedure to perform {\em posterior predictive checks} in the sense of \cite{rubin1984}. For two different electricity spot markets, over two different periods of time, we determine in this way the minimum number of superposed processes required in the model. We find that two or three factors are sufficient in each case. We also show that taking either constant or periodic deterministic jump intensity rates can provide a relatively simple but sufficiently flexible modelling palette. Since the jump processes influence the spot price directly (additively) and have their own proper dynamics such models are also rather interpretable. More generally, since multi-factor  models have been considered for a range of commodities including oil and gas \citep{schwartz2000short, brix2015estimation} our algorithm is also potentially applicable in these contexts although this is outside the scope of the present paper (see \citet[Chapter 5]{gonzalez_modelling_2015} for an application to gas prices).

Section \ref{sec:model} describes the model and the data which animates our study, while Section \ref{sec:inference} presents our MCMC algorithm including the approach to assessing model adequacy through posterior predictive checking. The data is analysed in  Section \ref{sec:real}. Section \ref{sec:discussion} contains discussion of results and Section \ref{sec:conclusion} concludes. Notes on the efficient implementation of the algorithm are provided in the appendix. Throughout we denote probability distributions as follows: $\mbox{N}(a,b)$ denotes the Normal distribution with mean $a$ and variance $b$, $\mbox{Ga}(a,b)$ the Gamma distribution with mean $a/b$, $\mbox{IG}(a,b)$ the Inverse-Gamma distribution with mean $b(a-1)^{-1}$ for $a>1$, $\mbox{Ex}(a)$ the Exponential distribution with mean $a$ and $\mbox{U}(a,b)$ the Uniform distribution on the interval $(a,b)$.

\section{Model}\label{sec:model}
\label{sec:motivation}
\begin{figure}[t]
\begin{centering}
\includegraphics[scale=0.7]{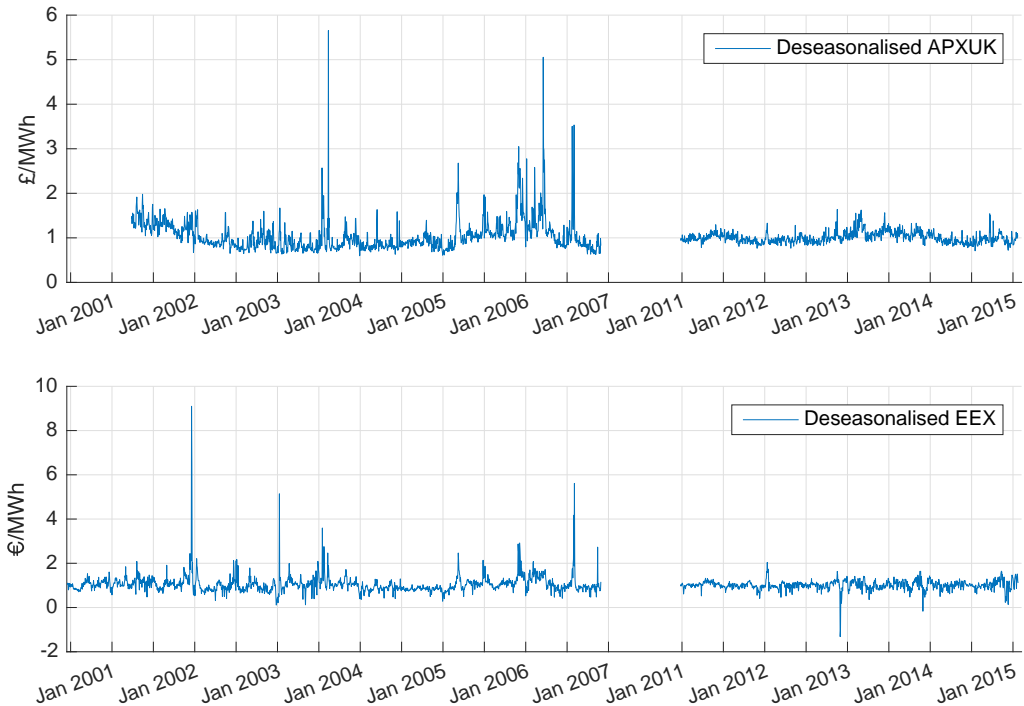}
\par\end{centering}
\caption{\label{fig:deseasonalised-data}
Deseasonalised APXUK (top panel) and EEX (bottom panel)  daily average prices (excluding weekends) over two periods. The first period starts on March 27, 2001 for APXUK and on June 16, 2000 for EEX and finishes on November 21, 2006 for both markets. The second period is January 24, 2011 to February 16, 2015 for both series. Details of the deseasonalisation procedure are given in the appendix.
}
\end{figure}

\subsection{Motivation}
Figure \ref{fig:deseasonalised-data} illustrates two electricity spot markets, the United Kingdom APXUK and European EEX, with weekend prices excluded. The left side of the figure plots daily average prices for the APXUK (March 2001 to November 2006) and the EEX (June 2000 to November 2006). This period was one of general growth in Europe, both economically and in electricity demand, and spot prices from this time have been studied by a number of authors including \citet{2203823}, \citet{meyer-brandis_multi-factor_2008} and \citet{benth_critical_2012}. The right hand side of the figure plots daily average price data between 2011 and 2015, a period of decline in UK electricity demand, although the picture across Europe was mixed.\footnote{Sources: European Commission Eurostat service; Digest of United Kingdom Energy Statistics (DUKES) 2015, UK Department of Energy \& Climate Change.} In order to reveal the structure of these price series more clearly the four time series have been separately deseasonalised (for details see the appendix). 

Reversion to a constant level is strongly suggested in the EEX data (bottom panel) and also, to a slightly lesser extent, in the APXUK series (top panel). Taking first the 2001--2006 APXUK data, the presence of significant positive price spikes is clear. However visual inspection also suggests that while some spikes decayed very quickly, a significant number showed more gradual decay. In contrast the positive spikes in the 2000--2006 EEX data appear uniformly to decay quickly and, in addition, the presence of smaller but rather frequent negative spikes is suggested. 

While the 2011--2015 APXUK data also suggests regular positive spikes, their heights are significantly smaller than those observed in 2001--2006. In the 2011--2015 EEX data the presence of negative spikes is suggested perhaps more strongly than in 2000--2006. Further, once these negative spikes are taken into account, visual inspection reveals apparently little evidence of positive spikes. 

For each series we apply the MCMC procedure described in Section \ref{sec:inference} to verify the conclusions of our visual analysis and to establish the smallest number of signed jump components for which the posterior predictive check is favourable, in a sense made precise in Section \ref{sec:real}.

In the following subsections we present in detail the class of spot price models to be calibrated.

\subsection{Ornstein-Uhlenbeck processes}\label{subsec:OU}

A process $Y(t)$, $0\leq t\leq T$ which is right continuous with left limits is called an
Ornstein-Uhlenbeck (OU) process if it is the unique strong solution to the stochastic differential
equation (SDE) 
\begin{equation}
dY(t)=\lambda^{-1}(\mu-Y(t))dt+\sigma dL(t),\quad Y(0-)=y_{0},\label{eq:OUprocess}
\end{equation}
where $L(t)$ is a driving noise process with independent increments, ie., a L\'{e}vy process. The initial state of the process $Y$ is defined as the value at the left-hand limit $Y(0-)$ due to the possibility of a jump at time $0$. In equation (\ref{eq:OUprocess}), $\mu\in\mathbb{R}$ is the mean
level to which the process tends to revert, $\lambda^{-1}>0$ denotes
the speed of mean reversion and $\sigma>0$ is the volatility of
the OU process.  The unique strong solution to the SDE (\ref{eq:OUprocess}) is
given by 
\begin{equation}
Y(t)=\mu + (y_{0} - \mu) e^{-\lambda^{-1}t}+\int_{0}^{t}\sigma e^{-\lambda^{-1}(t-s)}dL(s).\label{eq:sol-OU}
\end{equation}

We consider two different specifications for the L\'{e}vy process $L(t)$ driving $Y(t)$. On the one hand we consider an OU process where $L(t)=W(t)$
is a standard Wiener process. In this case the conditional distribution of 
$Y(t+s)$ given $Y(t)$, $t \in [0, T]$, $s \in [0,T-t]$, is Normal with the mean

\[
\mathbb{E}[Y(t+s)|Y(t)=y]=\mu+\left(y-\mu\right)e^{-\lambda^{-1}s},
\]
and the variance
\[
Var[Y(t+s)|Y(t)=y]=\lambda\sigma^{2}(1-e^{-2\lambda^{-1}s})/2.
\]
Hence we call the process $Y(t)$ a Gaussian OU process. On the other hand we consider the case where $L(t)$ is a compound Poisson process, with the {\em interval representation} 
\begin{equation}
L(t)=\sum_{j=1}^{\infty}\xi_{j}1_{\{t\geq\tau_{j}\}},\label{eq:interval-rep-for-L}
\end{equation}
where the $\tau_{j}$ are the arrival times of a Poisson process and $\xi_{j}$ represents the jump size at time $\tau_{j}$ (these jump sizes are independent and identically distributed (i.i.d.) random variables). The dynamics of $Y(t)$ are explicitly given by
\begin{equation}
Y(t+s)=\mu + (Y(t-) - \mu) e^{-\lambda^{-1}s}+\sum_{j: t\le\tau_{j}\leq t+s}e^{-\lambda^{-1}(t+s-\tau_{j})}\xi_{j}, \qquad s \geq 0.\label{eq:explicitY_Poisson}
\end{equation}
Below we shall model the stochastic part of energy spot prices by superimposing a number of OU processes.

\subsection{A multi-factor model for energy spot prices\label{sub:model-specification}}

Let us denote by $X(t)$ the de-trended and deseasonalised spot price at time $t\geq0$ (presentation of the relation between $X(t)$ and the electricity spot price $S(t)$ is deferred until the end of this section). We assume that the deseasonalised price $X(t)$ is a sum of $n+1$ OU processes 
\begin{equation}
X(t)=\sum_{i=0}^{n}w_{i}Y_{i}(t),\label{eqn:superposition_model}
\end{equation}
where $Y_{0}$ is a Gaussian OU process
\begin{eqnarray}
dY_{0}(t) & = & \lambda_{0}^{-1}(\mu-Y_{0}(t))dt+\sigma dW(t),\quad Y_{0}(0)=y_{0},\label{eq:Y0}
\end{eqnarray}
and each $Y_{i},i\geq1$ is a jump OU process
\begin{eqnarray}
dY_{i}(t) & = & -\lambda_{i}^{-1}Y_{i}(t)dt+dL_{i}(t),\quad Y_{i}(0-)=y_{i},i=1,\dots,n,\label{eq:Yi}
\end{eqnarray}
each $L_{i}$ being a (possibly inhomogeneous) compound Poisson process
with exponentially distributed jump sizes having mean $\beta_{i}$. We will refer to this as the {\em (n+1)-OU} model. The constants $w_{i}\in\{1,-1\}$ are used to indicate whether positive or negative jumps are being modelled. Notice that each of the processes $Y_{i}(t),i\geq1$, is non-negative since the $L_{i}$ are increasing processes. Thus by setting $w_{i}=1$, we employ $Y_{i}(t)$ to capture positive price spikes, whereas by setting $w_{i}=-1$,  $Y_{i}(t)$ is assumed to model negative price spikes. Throughout we assume that $w_0$ is equal to 1.

\begin{figure}
	\begin{centering}
	\includegraphics[scale=0.7]{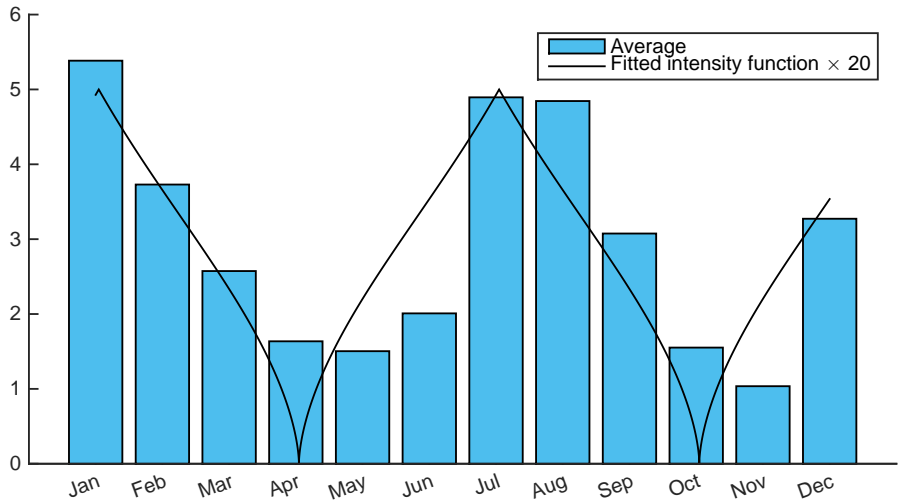} 
	\par\end{centering}
	\caption{\label{3c-EEX-average-number-pos-jumps}Monthly average number of positive jumps on the EEX market during 2000-6, together with the intensity function $I_1$ with parameter values taken from Table \ref{tab:prior-3OU-EEX-5Lup-1} for the 3-OU-$I_1$ model. As the jumps are not directly observable in the spot price series, we report the numbers inferred from the latent jump processes sampled in the MCMC procedure.}
\end{figure}

For each compound Poisson process $L_i,i\geq1$, we consider one of two specifications of the jump intensity rate. In the simpler specification we assume that the intensity rate is constant and equal to $\eta_i$, and hence that jump frequency is independent of time. In the alternative specification we take account of periodicity in the jump rate as in \citet{geman_understanding_2006} 
through the deterministic periodic intensity function
\begin{equation}
I_i(\eta_i,\theta_i,\delta_i,t) = \eta_{i}\left[\frac{2}{1+|\sin(\pi(t-\theta_{i})/k_i)|}-1\right]^{\delta_{i}},
\label{eq:periodic-intensity-rate}
\end{equation}
which has period $k_i$ days, where $k_i \in (0, \infty)$ (see Figure \ref{3c-EEX-average-number-pos-jumps} for a graph of the fitted intensity function $I_1$).
The parameter $\eta_i \in (0,\infty)$ 
is the maximum jump rate 
whilst the
exponent $\delta_i\in(0,\infty)$ controls the shape of 
the periodic function.
 In order to have a compact notation covering both the above model specifications, the intensity function parameter vector associated with the process $Y_i$ will simply be denoted $\vartheta_i$. In the constant intensity model specification we therefore understand that $\vartheta_i=\eta_i$, while in the periodic intensity model it is understood that $\vartheta_i=(\eta_i,\theta_i,\delta_i)$. 

We aim to show that using the sum of a number of such OU processes
provides suitable flexibility for modelling electricity spot prices. A diffusive Gaussian component is used to model regular trading characterised by frequent small price variations. The jump components model the arrival of temporary system disturbances of various kinds causing imbalance between supply and demand.
By specifying two jump components, say, such that
$\lambda_{1}>\lambda_{2}$, we can capture slowly and quickly decaying 
price spikes. As discussed in \cite{seifert_modelling_2007}, different decay rates may correspond to different {\em physical causes} of spikes such as power plant outages or extreme changes in weather. The possibility of incorporating negative price spikes
by taking $w_{i}=-1$  is explored in Section \ref{sec:real}.

In the empirical studies of Section \ref{sec:real} we assume that the relation between the spot price $S(t)$ and the deseasonalised price $X(t)$ is of the following form:
\begin{equation}
S(t)=e^{f(t/260)}X(t),\label{eq:Benth-spot}
\end{equation}
where $f:[0,\infty) \rightarrow \mathbb{R}$ is a deterministic function that captures the long-term price trend and seasonality typically observed in energy spot prices. In our analysis we take a day as the unit of time and skip weekends due to their distinctly different price dynamics, resulting in a 260-day year. The function $f$ is specified in terms of years to capture weather-induced market patterns linked to seasonal variations.
The multiplicative seasonality in \eqref{eq:Benth-spot} is in line with the exponential price trends standard in mathematical economics. We take
\begin{equation}
f(\tau; a_1, \ldots, a_6)=a_{1}+a_{2}\tau+a_{3}\sin(2\pi \tau)+a_{4}\cos(2\pi \tau)+a_{5}\sin(4\pi \tau)+a_{6}\cos(4\pi \tau),\label{eq:seasonal-trend}
\end{equation}
although our methodology applies to any other specification of seasonality provided its effect can be removed from the series of spot prices prior to statistical inference for $X(t)$.

\section{Inference} \label{sec:inference}

In this section we present a Markov Chain Monte Carlo (MCMC) approach to Bayesian inference in the superposition model (\ref{eqn:superposition_model}). We construct a Markov chain whose stationary distribution is the posterior distribution of the parameters in our model together with latent variables introduced to make the inference computationally tractable, see Section \ref{sub:Data-augmentation}. The application of a Gibbs sampler allows single parameters or groups thereof to be updated conditioned on others being fixed -- a standard MCMC approach which aids computational tractability. Central to the performance of MCMC and particularly the Gibbs sampler is the notion of {\em mixing} which is linked to the speed of convergence of the chain to its stationary distribution. Intuitively the better the mixing, the smaller the dependence between consecutive steps of the chain and, in effect, the less the chain gets blocked in small areas of the state space for long stretches of time. Mixing is negatively affected when the parameters which are updated at a given step of a Gibbs sampler depend on those upon which they are conditioned. This will be of particular importance in the choice of latent variables.

In Sections \ref{sub:Data-augmentation}-\ref{sec:MCMC_implementation} we present techniques for Bayesian inference in the superposition model (\ref{eqn:superposition_model}) in the case of one jump OU component ($n=1$). This is then extended in Section \ref{sec:Bayesian-inference-for-3-components}
to the case of multiple jump components.  For simplicity, when it does not lead to ambiguity, we drop the subscript in the jump process
$L_{1}$ and its parameters, so that $L=L_{1}$, $\beta=\beta_{1}$ and $\vartheta=\vartheta_{1}$. 

\subsection{Data augmentation\label{sub:Data-augmentation}}

Let $\obsX=\{x_{0},\dots,x_{N}\}$ denote observations of the process (\ref{eqn:superposition_model}) at times $0=t_{0},\dots,t_{N}=T$, and $\Delta_{i}=t_{i}-t_{i-1}>0$, $i=1,\dots,N$,
the time increments between consecutive observations. The likelihood $\ell(\obsX\mid\mu,\lambda_{0},\sigma,\lambda_1, \vartheta,\beta)$ of the data given parameters is neither analytically tractable nor amenable to numerical integration since it involves an infinite sum of integrals over high dimensional spaces. However by augmenting the state space with observations $\obsY_{1}=\{y_{1,0},\ldots,y_{1,N}\}$ of the process $Y_1$ at times  $t_{i}$,
the likelihood of $\obsX$ given $\obsY_1$ becomes independent of $\lambda_{1}$, $\vartheta$ and $\beta$. Thanks to the explicit form of the transition density of a Gaussian OU process we have
\begin{equation}
\ell(\obsX\mid\mu,\lambda_{0},\sigma,\obsY_{1})=\prod_{i=1}^{N}\frac{1}{\sqrt{2\pi}\Sigma_{i}}\exp\left\{ -\frac{1}{2\Sigma_{i}^{2}}\left(z_{i}-\mu-\left(z_{i-1}-\mu\right)e^{-\lambda_{0}^{-1}\Delta_{i}}\right)^{2}\right\} ,\label{eq:likelihood}
\end{equation}
where $\Sigma_{i}^{2}=\lambda_{0}\sigma^{2}(1-e^{-2\lambda_{0}^{-1}\Delta_{i}})/2$ and 
\begin{equation}
z_{i}=x_{i}-y_{1,i},\quad i=0,\dots,N.\label{eq:process-Z-bayesian-inference}
\end{equation}

Space augmentation methods have been widely used in statistics to tackle computationally infeasible problems. However the choice of latent variables or processes has a profound influence on the properties of the resulting estimators, affecting in particular the mixing of a Markov chain approximating the posterior distribution. From a mathematical point of view, the body of work closest to the present inference problem is estimation in the context of stochastic volatility models, where the volatility process is driven by a jump process. Among these are the state space augmentations used in \citep{jacquier_bayesian_1994,kim_stochastic_1998} and later criticised by
\citet[p. 188]{barndorff-nielsen_non-gaussian_2001} for high posterior correlation of the parameter $\lambda_{1}$ with the input trajectory of the process $Y_{1}$. This correlation could lead to MCMC samplers based on this parametrisation performing poorly. Instead, \citet{barndorff-nielsen_non-gaussian_2001} propose an alternative data augmentation scheme based
on a series representation of integrals with respect to a Poisson process $L(t)$,  known as the Rosi\'{n}ski or Ferguson-Klass representation. Bayesian inference for a stochastic volatility model under this parametrisation
was first explored in the discussion section of \citet{barndorff-nielsen_non-gaussian_2001} and further developed in \citet{griffin_inference_2006} and \cite{fruhwirth2009bayesian}. In the present paper, however, we opt for the following more direct parametrisation independently suggested by several researchers (see, for example, \citet{barndorff-nielsen_non-gaussian_2001} and  \citet{roberts_bayesian_2004}). 

Recall from Section \ref{sub:model-specification} that the process $L(t)$ driving $Y_1(t)$ is a compound Poisson process with intensity function $I(\vartheta,t)$ and interval representation \eqref{eq:interval-rep-for-L}. Recall also that
\begin{equation}
Y_1(t+s)=Y_1(t-)e^{-\lambda_1^{-1}s}+\sum_{j: t\le\tau_{j}\leq t+s}e^{-\lambda_1^{-1}(t+s-\tau_{j})}\xi_{j}, \qquad s \geq 0,\label{eq:explicitY2}
\end{equation}
which motivates a data augmentation methodology where the set of pairs $\{(\tau_{j},\xi_{j})\}$, instead of the process $Y$,  is treated as the missing data. This has the benefit of introducing independence between $\lambda_1$ and the latent variables, thus improving the mixing in Gibbs samplers.

Let us denote by $\Phi$ the marked Poisson process on $S=[0,T]\times (0,\infty)$
with locations $\tau_{i}$ on $[0,T]$ and marks $\xi_{i}$ on $(0,\infty)$.
The probability density of $\Phi$ is defined relative to a {\em dominating measure}, namely that of a Poisson process with unit intensity on $[0,T]$ and exponential jump sizes with parameter $1$. Hence, thanks to the marking theorem \citep{kingman_poisson_1992} and the likelihood ratio formula in \cite{kutoyants_statistical_1998}, the density of $\Phi$ with respect to this dominating measure is
\begin{equation} \label{eqn:density_of_Phi}
\ell(\Phi\mid\vartheta,\beta) = L(\vartheta;\Phi) \cdot \beta^{-N_T}\exp\Big\{-(\beta^{-1}-1)\sum_{j=1}^{N_T}\xi_{j}\Big\},
\end{equation}
where $N_T$ is the number of points in $S$. Here $L(\vartheta;\Phi)$ is the density, with respect to the Poisson process with unit intensity, of the Poisson process with intensity $I(\vartheta, t)$:
\begin{equation}\label{eqn:density_L}
L(\vartheta;\Phi) = \exp \left\lbrace \sum_{j=1}^{N_T} \log I(\vartheta,\tau_j)
				- \int_{0}^{T} I(\vartheta,t)dt + T 
	 \right\rbrace.
\end{equation}
When $L$ is a homogeneous Poisson process with constant intensity $\eta$ we obtain
\begin{equation}
L(\vartheta;\Phi)=\exp\big\{-(\eta-1)T\big\} \eta^{N_T}.\label{eq:Poisson-likelihood}
\end{equation}

We will use a Gibbs sampler  to simulate from the posterior distribution of the parameters and the missing data $\Phi$ given the observed data $\obsX$, using the factorisation
\begin{equation}
\pi(\mu,\lambda_{0},\sigma,\lambda_1,\vartheta,\beta,\Phi\mid \obsX)\propto\ell(\obsX\mid\mu,\lambda_{0},\sigma,\lambda_1,\Phi)\ell(\Phi\mid\vartheta,\beta)\pi(\mu,\lambda_{0},\sigma,\lambda_1,\vartheta,\beta),\label{eq:posterior}
\end{equation}
where $\pi(\mu,\lambda_{0},\sigma,\lambda_1,\vartheta,\beta)$ is the
joint prior density of the parameters.

\subsection{Classes of prior distributions}
\label{sec:priorspecification}
To complete our Bayesian model we now specify classes of prior
distributions for the parameters, which are assumed to be mutually independent\footnote{In Section \ref{sec:Bayesian-inference-for-3-components} and following, where more than one jump OU component is considered, the only statistical dependence we assume is a strict ordering of the mean reversion parameters $\lambda_{j}$, $j=1,\dots,n$ when this is needed for identifiability.}. For computational efficiency the classes chosen correspond to conjugate priors where possible.
In the empirical studies presented in Section \ref{sec:real} the prior distributions are chosen with a large spread (for example variance, where this exists) in order to {\em let the data speak for itself}. Prior expectations are based on existing results in the literature, combined with further exploratory analysis of historical data as necessary. For details see the appendix. Of course users of our methodology may also have prior beliefs about the model parameters, and in our Bayesian context the prior distributions  may alternatively be chosen to reflect these beliefs where appropriate.

 We specify a N$(a_{\mu},b_{\mu}^{2})$
prior distribution for the mean level $\mu$ of the Gaussian OU component, an $\mbox{IG}(a_{\sigma},b_{\sigma})$
for its volatility $\sigma^{2}$, an $\mbox{IG}(a_{\beta},b_{\beta})$ for the jump size parameter $\beta$ and an $\mbox{IG}(a_{\lambda_i},b_{\lambda_i})$
for the mean reversion parameter  $\lambda_i$, $i=0,1$. For the intensity function a $\mbox{Ga}(a_{\eta},b_{\eta})$ prior is chosen for $\eta$. Further when the intensity is periodic, a $\mbox{Ga}(a_{\delta},b_{\delta})$ prior is taken for $\delta$ and a $\mbox{U}(a_{\theta},b_{\theta})$ prior for $\theta$ (cf. \eqref{eq:periodic-intensity-rate}).

\subsection{MCMC algorithm}\label{sec:MCMC_implementation}
In the algorithm below the Gibbs step for updating the $\lambda_i$ employs a random-walk Metropolis-Hastings procedure. To ensure that the mixing is of the same order for small and large values of $\lambda_i$ the step of the proposal should be state dependent; equivalently an appropriate transformation of $\lambda_i$ may be applied. For computational convenience we opt for the latter, swapping $\lambda_i$ in the inference procedure with $\rho_i=e^{-\lambda_{i}^{-1}}$. 

After setting the initial state of the chain, the MCMC algorithm applied below cycles through the following steps:

\subsubsection*{MCMC algorithm for the 2-OU model}

\noindent Step 1: update $\mu \sim \pi(\mu\mid \rho_{0},\sigma, \rho_1, \obsX, \Phi)$\\
Step 2: update $\sigma^{2} \sim \pi(\sigma^2 \mid \rho_{0}, \rho_1, \obsX, \Phi)$\\
Step 3: update $\rho_{0}, \rho_1 \sim \pi(\rho_{0},\rho_1\mid \mu, \sigma, \obsX, \Phi)$\\
Step 4: update $\vartheta \sim \pi(\vartheta \mid \Phi)$ \\
Step 5: update $\beta \sim \pi (\beta \mid \Phi)$\\
Step 6: update $\Phi \sim \pi(\Phi \mid \mu,\rho_{0},\sigma,\rho_1, \vartheta, \beta, \obsX)$
\\
Step 7: Go to step 1.\\
\\
Below we provide more details about each of these steps.

\subsubsection*{Step 1. Update $\mu$}

Recalling \eqref{eq:likelihood}, the likelihood of the observed data conditional on the augmented state $(\mu,\lambda_{0},\linebreak\sigma,\rho_1,\Phi)$ is
\[
\ell(\obsX\mid\mu,\lambda_{0},\sigma,\rho_1,\Phi)
\propto \frac{1}{\prod_{i=1}^N\Sigma_i}\exp\left\{ -\frac{1}{2}\sum_{i=1}^{N}\frac{1}{\Sigma_i^{2}}\left(z_{i} - z_{i-1} e^{-\lambda_{0}^{-1}\Delta_i}+\mu\left(e^{-\lambda_{0}^{-1}\Delta_i}-1\right)\right)^{2}\right\},
\]
where $\Sigma_i^{2}=\lambda_{0}\sigma^{2}(1-e^{-2\lambda_{0}^{-1}\Delta_i})/2$ and the $z_i$ are computed as the difference between the observations of $X$ and the trajectory of $Y$ implied by the realisation $\Phi$ of the marked Poisson process. Using the conjugate prior for $\mu$ specified in the previous section it can be easily shown that the conditional distribution $\pi(\mu\mid \rho_0, \sigma,\rho_1, \obsX, \Phi)$ is
\[
\mbox{N}\left(
\frac{
\sum_{i=1}^{N}
\left(1-e^{-\lambda_{0}^{-1}\Delta_i}\right)\Sigma_i^{-2}
\left(z_{i}-z_{i-1}e^{-\lambda_{0}^{-1}\Delta_i}\right)
+\frac{a_{\mu}}{\sigma_{0}^{2}}
}
{\sum_{i=1}^N\left(1-e^{-\lambda_{0}^{-1}\Delta_i}\right)^{2}\Sigma_i^{-2}
+\frac{1}{b_{\mu}^{2}}},
\frac{1}{\sum_{i=1}^N\left(1-e^{-\lambda_{0}^{-1}\Delta_i}\right)^{2}\Sigma_i^{-2}
+\frac{1}{b_{\mu}^{2}}}\right).
\]

\subsubsection*{Step 2. Update $\sigma^{2}$}

Due to the choice of prior, the conditional distribution $\pi(\sigma^2 \mid \rho_{0}, \rho_1, \obsX, \Phi)$ has the closed form
\[
\mbox{IG}\left(
\frac{N}{2}+a_{\sigma},\frac{1}{\lambda_0}\sum_{i=1}^{N}\frac{s_i}{(1-e^{-2\lambda_{0}^{-1}\Delta_i})}
+b_{\sigma}\right
),
\]
where
\[
s_i=\left(z_{i}-z_{i-1}e^{-\lambda_{0}^{-1}\Delta_i}+\mu\left(e^{-\lambda_{0}^{-1}\Delta_i}-1\right)\right)^{2}.
\]

\subsubsection*{Step 3. Update $\rho_0$ and $\rho_1$}

Explicit conditional distributions for $\rho_0$ and $\rho_1$ are not available and the density is only known up to a multiplicative constant:
\begin{align*}
\pi(\rho_{0}\mid\mu,\sigma,\rho_1,\obsX,\Phi) &\propto \ell(\obsX\mid\mu,\rho_{0},\sigma,\rho_1, \Phi)\pi(\rho_{0}),\\
\pi(\rho_1\mid\mu,\sigma,\rho_{0},\obsX,\Phi)& \propto \ell(\obsX\mid\mu,\rho_{0},\sigma,\rho_1,\Phi)\pi(\rho_1).
\end{align*}
Hence we use a random-walk Metropolis-Hastings within Gibbs procedure to update $\rho_0$ and $\rho_1$.
The variance of the proposal distribution is tuned after pilot runs in order to achieve an acceptance rate between $20\%$ and $50\%$.

\subsubsection*{Step 4. Update $\vartheta$}

In the case of constant intensity function, the conjugate prior for $\eta$ yields an explicit conditional distribution
\begin{align*}
\eta \mid \Phi &\sim \text{Ga} \left(a_\eta + N_T, T + b_\eta \right).
\end{align*}
When the intensity function is time dependent we employ a random-walk Metropolis-Hastings within Gibbs procedure to update $\eta,\theta$ and $\delta$:
\[
\pi(\eta,\theta,\delta \mid \Phi) \propto \ell(\Phi\mid\eta,\theta,\delta) \pi(\eta)\pi(\theta) \pi(\delta).
\]
Here the non-explicit function $L(\vartheta |\Phi)$ of \eqref{eqn:density_L} is numerically calculated by a quadrature method. The variance of the proposal distribution is tuned after pilot runs as above.
\subsubsection*{Step 5. Update $\beta$}
The conditional distribution of $\beta$ given $\Phi$ has the closed form
\[
\beta\mid \Phi \sim \text{IG} \left(a_{\beta}+N_T,\sum_{i=1}^{N_T}\xi_{i}+b_{\beta}\right).
\]

\subsubsection*{Step 6. Update the latent process $\Phi$ }
\label{sec:phiupdate}

The Metropolis-Hastings step we use to update the process $\Phi$ draws from the work of \citet{geyer_simulation_1994}, \citet{roberts_bayesian_2004} and \citet{fruhwirth2009bayesian} on MCMC techniques for simulating point processes, extending it where appropriate to the case of inhomogeneous Poisson processes. 

Let us assume that the current state of the Markov chain is $$\Phi=\{(\tau_{1},\xi_{1}),\dots,(\tau_{N_T},\xi_{N_T})\},$$
that is, there are $N_T$ points on the set $S$ with jump times given by $\tau_{j}$ and the corresponding jump sizes by $\xi_{j}$. We choose randomly, with equal probability, one of the following three proposals.

\bigskip
\noindent\textit{Birth-and-death step}
\smallskip\nopagebreak

\noindent
In the birth-and-death step we choose one
of two moves. 
With probability $p\in(0,1)$ we choose
a birth move whereby a new-born point $(\tau,\xi)$ is added to the
current configuration of Poisson points. The proposed new state is
then $\Phi\cup\left\{ (\tau,\xi)\right\} $. The point $\tau$ is
drawn uniformly from $[0,T]$, whilst $\xi$ is drawn from the jump
size distribution $\mbox{Ex}(\beta)$. For this move the proposal
transition kernel $q(\Phi, \Phi \cup \{(\tau, \xi)\})$ has the following density with respect to the product of Lebesgue measure on $[0,T]$ and $\mbox{Ex}(1)$ measure on $(0, \infty)$:
\[
q(\Phi, \Phi \cup \{(\tau, \xi)\}) =\beta^{-1}\exp\left(-(\beta^{-1}-1)\xi\right).
\]
With probability $1-p$ a death move is selected, a randomly chosen point $(\tau_i,\xi_i)$ being removed from $\Phi$ (provided that $\Phi$ is not empty).
The proposal transition kernel (with respect to the counting measure) is
\[
q(\Phi, \Phi \setminus \{(\tau_i, \xi_i)\}) = \frac{1}{N_T},
\]
where $N_T$ is the number of points in $\Phi$ before the death move. Then the Metropolis-Hastings acceptance ratio for a birth move from $\Phi$ to $\Phi\cup\{(\tau,\xi)\}$ is
\[
\alpha(\Phi,\Phi\cup\{(\tau,\xi)\})=\min\left\{ 1,r(\Phi,(\tau,\xi))\right\} ,
\]
while the acceptance ratio for a death move from $\Phi$ to $\Phi \setminus \{(\tau_i,\xi_i)\}$ is
\[
\alpha(\Phi,\Phi\setminus\{(\tau_i,\xi_i)\})=\min\left\{ 1,\frac1{r\big(\Phi \setminus \{(\tau_i,\xi_i)\},(\tau_i,\xi_i)\big)}\right\},
\]
 where
 \begin{eqnarray*}
 	r(\tilde \Phi,(\tau,\xi))
 	& = & \frac{\ell(\obsX\mid\mu,\rho_{0},\sigma,\rho_1,\tilde \Phi  \cup \{ (\tau, \xi)\} )}
			   {\ell(\obsX\mid\mu,\rho_{0},\sigma,\rho_1,\tilde \Phi)}
 	\frac{\pi(\tilde \Phi  \cup \{ (\tau, \xi)\}\mid \vartheta,\beta )}
		 {\pi(\tilde \Phi\mid \vartheta,\beta)}
 	\frac{1-p}{p} \\
 	&&\times \frac{1}{(N_T+1)q(\Phi, \Phi \cup \{(\tau, \xi)\})}\\
 	& = & \frac{\ell(\obsX\mid\mu,\rho_0,\sigma,\rho_1,\tilde \Phi  \cup \{ (\tau, \xi)\})}
 	           {\ell(\obsX\mid\mu,\rho_0,\sigma,\rho_1,\tilde \Phi)}  \frac{1-p}{p}\frac{T}{\tilde N_T+1} I(\vartheta,\tau),
 \end{eqnarray*}
where $\tilde N_T$ is the number of points of $\tilde \Phi$, cf. \eqref{eqn:density_of_Phi}.

\bigskip
\noindent\textit{Local displacement move}
\smallskip\nopagebreak

\noindent
Without loss of generality let us assume that the
jump times of the Poisson process are ordered, so that $\tau_{1}<\dots<\tau_{N_T}$. In the local displacement move we choose randomly
one of the jump times, say $\tau_{j}$, and generate a new jump time
$\tau$ uniformly on $[\tau_{j-1},\tau_{j+1}]$, putting
 $\tau_{0}=0$ and $\tau_{N_T+1}=T$. The point $(\tau_{j},\xi_{j})$ is then displaced and re-sized to $(\tau, \xi)$, where $\xi = e^{-\lambda_1^{-1}(\tau-\tau_{j})}\xi_{j}$. Formally we choose uniformly one of $N_T$ transition kernels,  with the $j$-th one preserving the conditional distribution $\pi(\tau,\xi|\obsX,\mu,\rho_{0},\sigma,\rho_1,\vartheta,\beta,\Phi\backslash\{(\tau_{j},\xi_{j})\})$. The proposal for the $j$-th kernel has the Uniform distribution over $(\tau_{j-1}, \tau_{j+1})$ for the first variable with the second variable being then a deterministic transformation given by a 1-1 mapping $\mathcal{T}(\xi, \tau, \tau') = (\xi e^{-\lambda_1^{-1}(\tau'-\tau)}, \tau', \tau)$ such that $\mathcal{T} = \mathcal{T}^{-1}$. Following \citet[Section 2]{tierney_note_1998}, the contribution of this deterministic transition to the Metropolis-Hastings acceptance ratio is $|\det \nabla \mathcal{T} (\xi_j, \tau_j, \tau) |$, where $\tau$ is the new proposed location of the jump. Hence
the complete Metropolis-Hastings acceptance ratio is
\begin{multline*}
r(\Phi, \Phi_{new}) = \frac{\ell(\obsX|\mu,\rho_{0},\sigma,\rho_1,\Phi_{new})}{\ell(\obsX|\mu,\rho_{0},\sigma,\rho_1,\Phi)}\frac{\pi(\tau,\xi|\vartheta,\beta)}{\pi(\tau_{j},\xi_{j}|\vartheta,\beta)}\frac{\tilde{q}(\tau,\tau_{j})}{\tilde{q}(\tau_{j},\tau)}|\det \nabla \mathcal{T} (\xi_j, \tau_j, \tau) |\\
=
\frac{\ell(\obsX|\mu,\rho_{0},\sigma,\rho_1,\Phi_{new})}{\ell(\obsX|\mu,\rho_{0},\sigma,\rho_1,\Phi)}\frac{I(\vartheta,\tau)}{I(\vartheta,\tau_{j})}\frac{e^{-\beta^{-1}\xi}}{e^{-\beta^{-1}\xi_{j}}}e^{-\lambda_{1}^{-1}(\tau-\tau_{j})},
\end{multline*}
where $\tilde q(\tau, \tau') = (\tau_{j+1}- \tau_{j-1})^{-1}$ is the transition density for the jump location with respect to Lebesgue measure on $(\tau_{j-1}, \tau_{j+1})$.

\bigskip
\noindent\textit{Multiplicative jump size update}
\smallskip\nopagebreak

\noindent
In this step the sizes of all jumps are independently updated. Specifically, for each jump $(\tau_j, \xi_j)$ we propose a new jump size $\xi_{j}'=\xi_{j}\phi_{j}$, where $\log(\phi_j) \sim \mbox{N}(0,c^{2})$ are i.i.d. random variables.
The variance $c^2$ is chosen inversely proportional to the current number of jumps, and the performance of this update step appears rather insensitive to the constant of proportionality. Denoting by $\Phi_{new}$ the Poisson point process with updated jump sizes, the Metropolis-Hastings acceptance ratio for this move is
\begin{align*}
\alpha(\Phi,\Phi_{new})
= \min\left\{ 1,
	\frac{\ell(\obsX\mid\mu,\rho_0,\sigma,\rho_{1},\vartheta,\beta,\Phi_{new})}
		    {\ell(\obsX\mid\mu,\rho_0,\sigma,\rho_{1},\vartheta,\beta,\Phi)}
		    \exp\left\lbrace-(\beta^{-1}-1)\sum_{i=1}^{N_T}(\xi_{i}'-\xi_{i})\right\rbrace\prod_{i=1}^{N_T}\frac{\xi_{i}^{'}}{\xi_{i}}
		    \right\}.
\end{align*}
Note that the product $\prod_{i=1}^{N_T}\frac{\xi_{i}^{'}}{\xi_{i}}$ is equivalent to $\prod_{i=1}^{N_T}\phi_{i}$.

\subsection{Bayesian inference for a sum of three OU processes}\label{sec:Bayesian-inference-for-3-components}

As discussed in Section \ref{sub:model-specification} we may believe {\em a priori} that the price spikes observed in the market are of a certain number of types, corresponding to their differing possible physical causes. Accordingly we now describe the extension of the 2-OU model by the addition of a further independent jump OU component $Y_{2}(t)$, henceforth referring to the first jump component as $Y_{1}(t)$ and using appropriate subscripts to distinguish their parameters; further jump components are incorporated similarly. The new jump component $Y_2$ may either have a positive contribution to the price process (ie. the sign $w_2$ in \eqref{eqn:superposition_model} is $1$) with a rate of decay differing from that of the first component, or alternatively it may have a negative contribution (ie. $w_2=-1$). For concreteness here we choose $w_1=w_2=1$, so that:
\begin{equation}
	X(t)=Y_{0}(t)+Y_{1}(t)+Y_{2}(t),\label{eq:3component-model}
\end{equation}
and we specify that $\lambda_{1}>\lambda_{2}$ for identification purposes, ie. the jumps of $Y_{1}(t)$ have slower decay than those of $Y_{2}(t)$. When $w_1=1, w_2=-1$ this constraint is not required.

We now have two marked Poisson processes $\Phi_{1}$ and $\Phi_{2}$, corresponding to $L_{1}(t)$
and $L_{2}(t)$ respectively, which are conditionally independent given their parameters. 
The augmented likelihood $\ell(\obsX \mid\mu,\lambda_{0},\sigma,\obsY_{1}, \obsY_{2})$
is given by equation (\ref{eq:likelihood})  with $z_{j}=x_{j}-y_{1,j}-y_{2,j}$.
The likelihood of $\Phi=(\Phi_{1},\Phi_{2})$
with respect to the dominating measure of a pair of independent marked Poisson process with unit intensity and jump sizes with a Ex$(1)$ distribution is given by 
\[
\pi(\Phi\mid\vartheta_{1},\vartheta_{2},\beta_{1},\beta_{2})=\pi(\Phi_{1}\mid\vartheta_{1},\beta_{1})\pi(\Phi_{2}\mid\vartheta_{2},\beta_{2}),
\]
where for $i=1,2$, 
\[
\pi(\Phi_{i}\mid\vartheta_{i},\beta_{i}) =
	\exp \left\lbrace \sum_{j=1}^{N_T^i} \log I_i(\vartheta_i,\tau_{i,j})
- \int_{0}^{T} I_i(\vartheta_i,t)dt + T 
\right\rbrace 
\beta_i^{N_T^i}
\exp\left\{ -(\beta_{i}^{-1}-1)\sum_{j=1}^{N_T^i}\xi_{i,j}\right\}. 
\]
Here $N_T^i$ denotes the number of points of $\Phi_{i}$ in
the set $S$ and 
\[
\Phi_{i}=\{(\tau_{i,1},\xi_{i,1}),\ldots,(\tau_{i,N_T^i},\xi_{i,N_T^i})\}.
\]
We impose the condition $\lambda_1>\lambda_2$ by defining the prior distribution for $\rho_2$ as 
\[
\rho_{2}|\rho_{1}\sim\rho_{1}\mbox{U}(0,1),
\]
where $\rho_{i}=e^{-1/\lambda_{i}}, i\geq 1$, as before (see the appendix for properties of $\rho_2$). Since all other parameters are {\em a priori} mutually independent, their update steps are identical to those for the 2-OU model.

Updates of $\rho_{1}$ and $\rho_{2}$ are made using a random-walk Metropolis-Hastings algorithm (with independent proposals  for each variable). The proposal distribution is Normal with the variance tuned as before after pilot runs.

\subsection{Posterior predictive check} \label{sec:Diagnostics-theory}

Our approach to assessing model adequacy is the following. Since $Y_{0}$ is a Gaussian OU process, its transition density specified in Section \ref{subsec:OU} implies that $\varepsilon_j$, $j=1, \ldots, N,$ defined implicitly by 
\begin{equation}
Y_{0}(t_j)=\mu+(Y_{0}(t_{j-1})-\mu)e^{-\lambda_{0}^{-1}\Delta_j}+\left(\frac{\sigma^{2}\lambda_{0}}{2}(1-e^{-2\lambda_{0}^{-1}\Delta_j})\right)^{1/2}\varepsilon_{j},\label{eq:Explicit-OU-process}
\end{equation}
are independent and distributed as $\mbox{N}(0,1)$. Given observations of $Y_{0}$ at sampling times $t_0, \ldots, t_N$ the distribution of $\{\varepsilon_{1},\dots,\varepsilon_{N}\}$ may then be tested, and this test may be repeated across MCMC iterations. At each iteration $k$ of the MCMC algorithm (assuming the Markov chain has reached stationarity) we use the current state of parameters  $\Theta^{(k)}=\{\mu^{(k)},\lambda_{i}^{(k)},\sigma^{(k)},\vartheta_{i}^{(k)},\beta_{i}^{(k)}\}$ and missing data $\Phi^{(k)}$ to recover the path of each jump process $y_{i,j}^{(k)} ,j=0,\dots,N$. The path of $Y_{0}^{(k)}$ at times $t_0, \ldots, t_N$ is then computed as 
\begin{equation}
z^{(k)}_{j} = x_{j}-\sum_{i=1}^{n} w_i y_{i,j}^{(k)}, \quad j=0, \ldots, N, \label{eq:KS-Y0k}
\end{equation}
where $x_{i}$ is the deseasonalised price at time $t_i$ and $w_i$ is the sign of the $i$-th jump component.  From this the noise data $\{\varepsilon_{j}^{(k)}\}_{j=1,\ldots,N}$ for each MCMC iteration $k$ is obtained and subjected to a Kolmogorov-Smirnov (KS) test for the standard Normal distribution, yielding a $p$-value $p^{(k)}$.
Following \citet{rubin1984} we call the distribution of $p^{(k)}$ the posterior predictive check distribution. We refer to its mean as the {\em posterior predictive $p$-value} and interpret it accordingly, cf. \citet{gelman_bayesian_2003} and references therein.

Similarly we perform diagnostics on the jump processes sampled from our Markov chain. For each jump process $L_i$ at iteration $k$ of the Markov chain, the set of jump sizes and the set of inter-arrival times are both subjected to KS tests. The former set undergoes a KS test for the Exponential distribution with mean $\beta_i^{(k)}$. In the homogeneous Poisson process model, the latter set undergoes a KS test for the Exponential distribution with mean $(\eta_i^{(k)})^{-1}$; otherwise an independent sample is taken from the inter-arrival times of an inhomogeneous Poisson process with time varying intensity $I_i(\vartheta^{(k)}_i,t)$ and its distribution compared with the latter set in 
 a two-sample KS test. Posterior predictive $p$-values are reported.
 
\subsection{Implementation}

The parameter-dependent balance between jumps and diffusion in the spot price model \eqref{eqn:superposition_model} raises a number of potential issues regarding the implementation of the MCMC procedure described above. Extensive testing was carried out with simulated data in order to probe these issues, and details are given in the appendix.

Our MCMC algorithms were implemented by combining Matlab and C++ MEX code (provided as an electronic supplement), and were run on a 2.5GHz Intel Xeon E5 processor. For the 2-OU model with $1500$ observations  the computation time for completing 1000 MCMC iterations using a single core was approximately 1.9 seconds. The corresponding figure for the 3-OU model was about 3.6 seconds with a single update of the latent process $\Phi$, and 8.8 seconds with 5 updates of $\Phi$ per MCMC iteration. In the numerical examples of Section \ref{sec:real}, a burn-in period of 500\,000 iterations was allocated. The following 1.5 million were thinned by taking one sample every 100 iterations and used to establish the posterior distribution. This corresponds to around 1 hour running time for the 2-OU model and below 5 hours for the 3-OU model with the fivefold update of the latent process $\Phi$.

\section{Case study application to the APXUK and EEX markets}\label{sec:real}
In this section we apply the inference procedure described in Section \ref{sec:inference} to daily average electricity prices corresponding to the APX Power UK spot base index (APXUK hereafter, quoted in \pounds/MWh)%
\footnote{\url{https://www.apxgroup.com/market-results/apx-power-uk/ukpx-rpd-index-methodology/}%
} and the European Energy Exchange Phelix day base index (EEX, quoted in \euro/MWh)%
\footnote{\url{http://www.epexspot.com/en/}%
}. 
The data were retrieved from Thompson Reuters Datastream. Weekends (which tend to differ statistically from weekdays due to a reduction in trading) are removed in both cases. Hence we assume a calendar year of 260 days and take $\Delta_j=1$ so that parameters are reported in daily units.
We consider two time periods, namely 2000--2006 and 2011--2015. Figure \ref{fig:deseasonalised-data} displays deseasonalised versions of this data. The pattern of positive and negative spikes appears to differ across the two markets and time periods and correspondingly we perform four separate analyses.

A step-by-step guide to the practical application of our techniques is first presented in Section \ref{subsec:stepbystep}, followed in Section \ref{subsec:summ} by an economics-oriented discussion of the models fitted to the above datasets. The steps involved in fitting  are collected in Sections \ref{subsec:maths}--\ref{subsec:2011-15}, together with related econometric discussion.

\subsection{Step-by-step guide}\label{subsec:stepbystep}

This section contains a subjective guide to estimation for the above multi-factor price model using discrete observations. 
The analytical procedure is summarised in an algorithm below and later exemplified in Sections \ref{subsec:maths}-\ref{subsec:2011-15}. 
 
\begin{enumerate}
\item Deseasonalise time series
\item Set $n=1$
\item Fit all combinations of $(n+1)$-OU models (by a combination we mean the number of positive and negative jump components) \label{step3}
\item For each combination in step \ref{step3}:
\begin{enumerate}
\item compute posterior predictive $p$-values for the increments of the residual Gaussian OU process,
\item for each latent Poisson process, compute posterior predictive $p$-values for the jump arrival rates and the jump sizes
\end{enumerate}
\item \label{step5} Accept a model if all $p$-values are above a selected threshold (we have used $10\%$)
\item If no model has been accepted then, for each combination in step \ref{step3}:
introduce time-varying intensities for every subset of the jump components (see Eq.\eqref{eq:periodic-intensity-rate}); repeat model estimation, compute predictive $p$-values and accept / reject model as in step \ref{step5}.
\item If no model has been accepted then set $n:=n+1$ and go back to step 3.
\end{enumerate}

The above procedure suffers from the classical problem of multiple testing \citep{Benjamini1995}, ie., if sufficiently many models are tested one of them will prove statistically significant purely due to randomness. However, for the sake of simplicity we decided not to include this aspect of model selection in the above procedure. Instead we suggest verifying the selected model on a subset of the data, or on another dataset with similar characteristics, in order to confirm that the model choice is robust. This involves estimating the selected model on the new dataset and checking that all predictive $p$-values remain above the threshold applied.

The appropriate modelling of the seasonal component, in order to remove it from the data, depends heavily on the market under study \citep{weron_modeling_2007}. 
Electricity spot markets experience significant yearly variations and the specification \eqref{eq:seasonal-trend} of the seasonality function takes these into account. We used a minimum least-squares fit for the log price, which corresponds to a linear regression of the natural logarithm of price on each of the terms of the seasonality function $f$.

According to the algorithm above, the data is first deseasonalised. It is then checked whether the simplest model with just one jump component provides a suitable statistical description of the data, as follows. The  MCMC procedure is applied both to the model with one positive jump component and to the model with one negative jump component. For each model assessed this generates a sequence of posterior samples of: the latent Poisson process driving the price spikes; the spike sizes; and the implied discrete increments of the latent Gaussian Ornstein-Uhlenbeck process. These are used to compute posterior predictive $p$-values for the jump times, jump sizes and the increments of the Gaussian OU process as in Section \ref{sec:Diagnostics-theory}. 
If all computed $p$-values lie above a predetermined threshold (we have used $10\%$) the model may be deemed statistically valid. If neither of the simple models is statistically valid, we suggest relaxing the assumption of constant jump intensity and repeating the estimation procedure. If those models also fail the statistical validity test, the number of jump components may be increased by one and the above estimation and verification procedure repeated. Since the inclusion of further jump components clearly improves the model fit, this motivates the acceptance of the simplest model satisfying the above criterion.

\subsection{Summary of empirical results}\label{subsec:summ}

Our first hypothesis in this work is that, over time, disturbances in the different drivers involved in electricity spot price formation give rise to spikes with statistically distinguishable directions, frequencies, height distributions and rates of decay. Our results provide evidence for this hypothesis in data from the period 2000--2006. For the 2001--2006 APXUK data we find that the model with a single positive jump component has posterior predictive $p$-values which are too low to be judged adequate. In contrast when two independent positive jump components are included in the model, the MCMC procedure is able to distinguish these two factors statistically. This is evidenced by posterior predictive $p$-values for each fitted component which are statistically acceptable.

Similarly in the 2000--2006 EEX data our calibration procedure is able to statistically distinguish two independent jump components. In this case it is the signed combination of one positive and one negative jump component which has empirical support (that is, acceptable posterior predictive $p$-values for each fitted component). Both the model with a single positive jump component, and the model specifying two positive jump components, have posterior predictive $p$-values which are too low to be judged adequate.

Our second hypothesis is that the composition of these statistical models evolves in parallel with changes in underlying economic factors. This hypothesis is supported by comparing the above models for 2000--2006 with models for the same markets over the period 2011--2015. In both cases statistical changes are detected and, further, the nature of the change is consistent across the two markets. In this later period   a single positive jump component provides an adequate fit to the APXUK data. There is also an apparent decrease in the frequency of price spikes: on average the total number of jumps (of any size) per unit time is less than a third for the 2011--2015 series compared to 2001--2006. The 2011--2015 EEX data also has one fewer positive jump component relative to the period 2000--2006. Thus in the later period a single negative jump component is adequate.

\subsection{Deseasonalisation of the time series}\label{subsec:maths}

In this paper we perform inference on deseasonalised data, treating the seasonal trend function $f(t)$ as a known characteristic of the particular energy market under study. For the purposes of the numerical illustration in this section we assume the form \eqref{eq:seasonal-trend} and solve a least squares problem 
\[
\sum_{i=0}^{N}\left(\log S_{obs}(t_i)-f(t_i/260)\right)^{2} \to \min,
\]
where $S_{obs}(t)$ denotes the observed spot price at time $t$. Table A.3 in the appendix presents estimated parameters for the trend functions and
Figure \ref{fig:deseasonalised-data} displays the resulting deseasonalised time series $X(t)=S_{obs}(t)e^{-f(t/260)}$. 

\subsection{2001--2006 APXUK data}

\begin{table}[tb]
\begin{onehalfspace}

\begin{center}
	\begin{tabular}{cccccccc}
		\hline 
		&  & \multicolumn{3}{c}{Prior properties } &  & \multicolumn{2}{c}{Posterior properties}\tabularnewline
		\cline{3-8} 
		Parameter &  & Prior & Mean & SD &  & Mean & SD\tabularnewline
		\hline 
		$\mu$ &  & $\mbox{N}(1,20^{2})$ & 1 & 20 &  & 0.9592 & 0.0308\tabularnewline
		$\sigma^{2}$ &  & $\mbox{IG}(1.5,0.005)$ & 0.01 & - - &  & 0.0096 & 0.0008\tabularnewline
		$e^{-1/\lambda_{0}}$ &  & $\mbox{U}(0,1)$ & 0.5 & $0.2887$ &  & 0.9170 & 0.0116\tabularnewline
		$(\lambda_{0})$ &  & {(\mbox{IG}(1,1))} & (- -)  & (- -) &  & (11.7936) & (1.8298)\tabularnewline
		$e{}^{-1/\lambda_1}$ &  & U$(0,1)$ & 0.5 & $0.2887$ &  & 0.1570 & 0.0189\tabularnewline
		$(\lambda_1)$ &  & {(\mbox{IG}(1,1))} & (- -) & (- -) &  & (0.5403) & (0.0352)\tabularnewline
		$\eta$ &  & $\mbox{Ga}(1,10)$ & 0.1 & 0.1 &  & 0.2499 & 0.0297\tabularnewline
		$\beta$ &  & $\mbox{IG}(1,1)$ & - - & - -  &  & 0.7159 & 0.0738\tabularnewline
		\hline 
	\end{tabular}
	\par\end{center}
\end{onehalfspace}

\caption{\label{tab:2OU-APXUK-prior-and-estimates} Prior distributions and posterior moments
obtained when calibrating the 2-OU model to the 2001--2006 APXUK data.  The posteriors for the `indirect' parameters $\lambda_i$ were obtained by transformation of the parameters $e{}^{-1/\lambda_i}$ of the Markov chain at each step and their entries are given in brackets. }
\end{table}

\subsubsection{One jump component}\label{sec:APXsingle}
We take the priors specified in Table \ref{tab:2OU-APXUK-prior-and-estimates} as input to the 2-OU model with a single, positive jump component and constant intensity rate. The Markov chain was initialised with the state $(\mu,\lambda_{0},\sigma,\lambda_{1},\eta,\beta, \Phi)=(1,5,0.1,2,0.1,0.5,\mathbf{0})$ where $\mathbf{0}$ denotes the  absence of jumps, and the birth-and-death parameter $p$ was set equal to 0.5 (ie. equal probability for the birth or death of a jump). Table \ref{tab:2OU-APXUK-prior-and-estimates} presents summary statistics for the prior and posterior distributions of the univariate parameters. For those priors with a second moment, it may be observed that the standard deviation of the posterior is typically at least an order of magnitude smaller.

\subsubsection{Two jump components}

We also calibrate a 3-OU model with constant jump intensity and two positive jump components with the condition $\lambda_1 > \lambda_2$ introduced for identifiability (through the use of an appropriate prior as specified in Subsection \ref{sec:Bayesian-inference-for-3-components}), ie. the jumps of $Y_2$ decay faster than those of $Y_1$. The initial state of the chain was set to $$(\mu,\lambda_{0},\sigma,\lambda_{1},\eta_{1},\beta_{1},\lambda_{2},\eta_{2},\beta_{2},\Phi)=(1,5,0.2,5,0.001,0.5,1,0.001,0.5, \mathbf{0}).$$ 
Summary statistics for the prior and posterior distributions of the univariate parameters are given in Table \ref{tab:prior-3OU-APXUK}. For each of the jump component parameters $\lambda_i$, $\eta_i$ and $\beta_i$ the posterior distributions for the two jump components $i=1,2$ are well separated. 
In particular the `new' jump component $Y_{2}$ suggests that the quickly decaying price shocks are both less frequent and larger on average than the more slowly decaying jumps given by $Y_1$. As may be anticipated, the posterior distribution of the volatility $\sigma$ of the diffusion component $Y_{0}$ is correspondingly shifted lower with the inclusion of $Y_2$. However the posterior moments of the speed of mean reversion $\lambda_0$ remained virtually unchanged.

\begin{table}[ht]
	\begin{onehalfspace}
		\begin{center}
			\begin{tabular}{cccccccc}
				\hline 
				&  & \multicolumn{3}{c}{Prior properties } &  & \multicolumn{2}{c}{Posterior properties}\tabularnewline
				\cline{3-8} 
				Parameter &  & Prior & Mean & SD &  & Mean & SD\tabularnewline
				\hline 
				$\mu$ &  & $\mbox{N}(1,20^{2})$ & 1 & 20 &  & 0.8693 & 0.0284\tabularnewline
				$\sigma^{2}$ &  & $\mbox{IG}(1.5,0.005)$ & 0.01 & - - &  & 0.0057 & 0.0006\tabularnewline
				$e^{-1/\lambda_{0}}$ &  & $\mbox{U}(0,1)$ & 0.5 & 0.2887 &  & 0.9176 & 0.0128\tabularnewline
				$(\lambda_{0})$ &  & ($\mbox{IG}(1,1)$)  & (- -) & (- -) &  & (11.9352) & (1.9892)\tabularnewline
				$e{}^{-1/\lambda_{1}}$ &  & U$(0,1)$ & 0.5 & 0.2887 &  & 0.6605 & 0.0342\tabularnewline
				$(\lambda_{1})$ &  & ($\mbox{IG}(1,1)$) & (- -) & (- -) &  & (2.4408) & (0.3075)\tabularnewline
				$e^{-1/\lambda_{2}}$ &  & - - & 0.25 & 0.2205 &  & 0.0742 & 0.0148\tabularnewline
				$(\lambda_{2})$ &  & (- -) & (- -) & (- -) &  & (0.3839) & (0.0297)\tabularnewline
				$\eta_{1}$ &  & $\mbox{Ga}(1,10)$ & 0.1 & 0.1 &  & 0.2412 & 0.0487\tabularnewline
				$\eta_{2}$ &  & $\mbox{Ga}(1,10)$ & 0.1 & 0.1 &  & 0.1698 & 0.0250\tabularnewline
				$\beta_{1}$ &  & $\mbox{IG}(1,1)$ & - - & - - &  & 0.2243 & 0.0291\tabularnewline
				$\beta_{2}$ &  & $\mbox{IG}(1,1)$ & - - & - - &  & 0.8544 & 0.1047\tabularnewline
				\hline 
			\end{tabular}
			\par\end{center}
	\end{onehalfspace}
	
	\caption{\label{tab:prior-3OU-APXUK}Prior distributions and posterior properties obtained when calibrating the 3-OU model to the 2001--2006 APXUK data. The indirect parameters $\lambda_i$ are treated as described in the caption to Table \ref{tab:2OU-APXUK-prior-and-estimates}. The distribution and moments of $\rho_2=e^{-1/\lambda_2}$ are calculated in the elecronic appendix.}
\end{table}

\subsubsection{Augmentation of the state space}\label{sub:model-comparison-APXUK}
\begin{figure}[p]
	\begin{centering}
		\includegraphics[scale = 0.7]{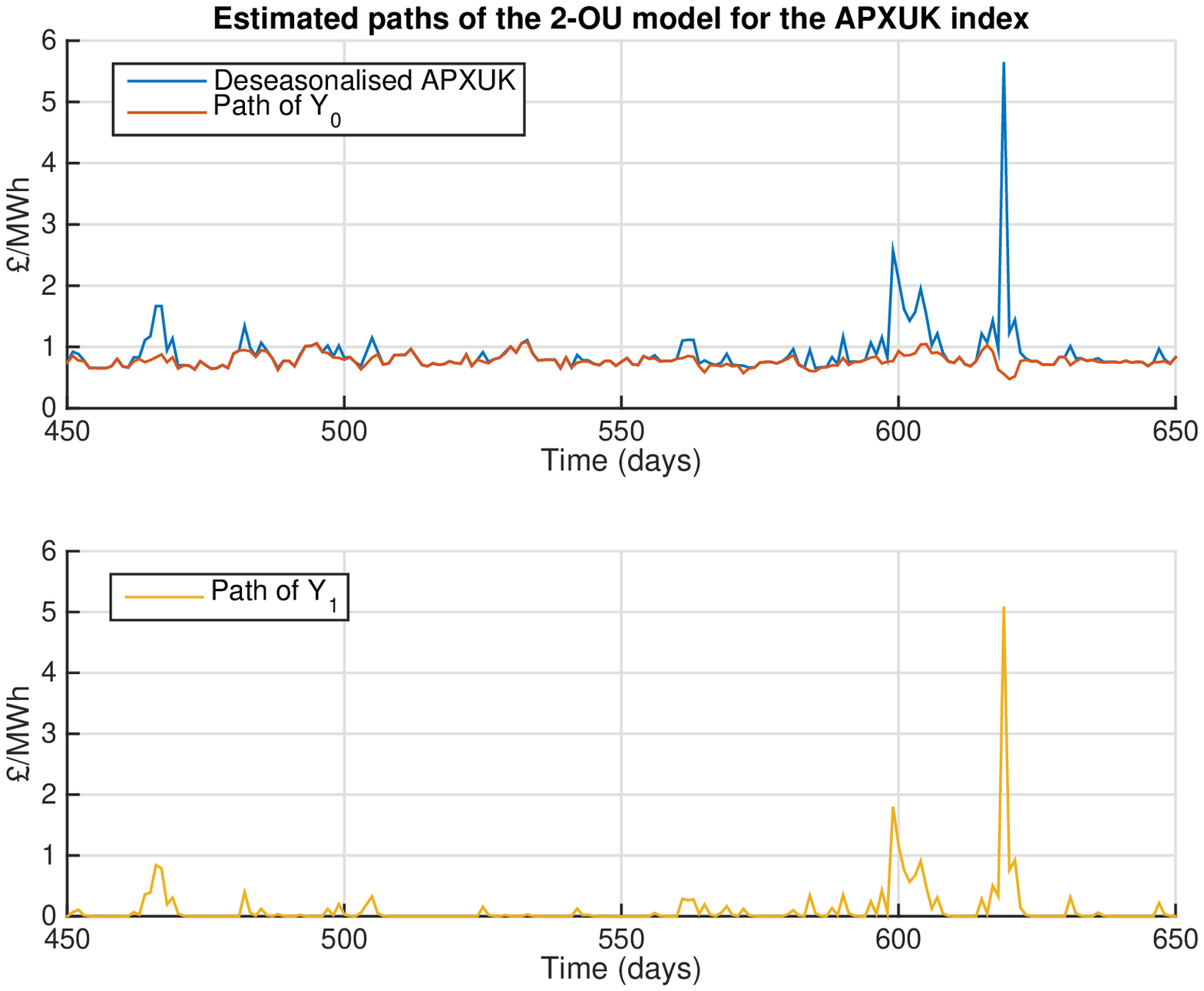}
		\includegraphics[scale = 0.7]{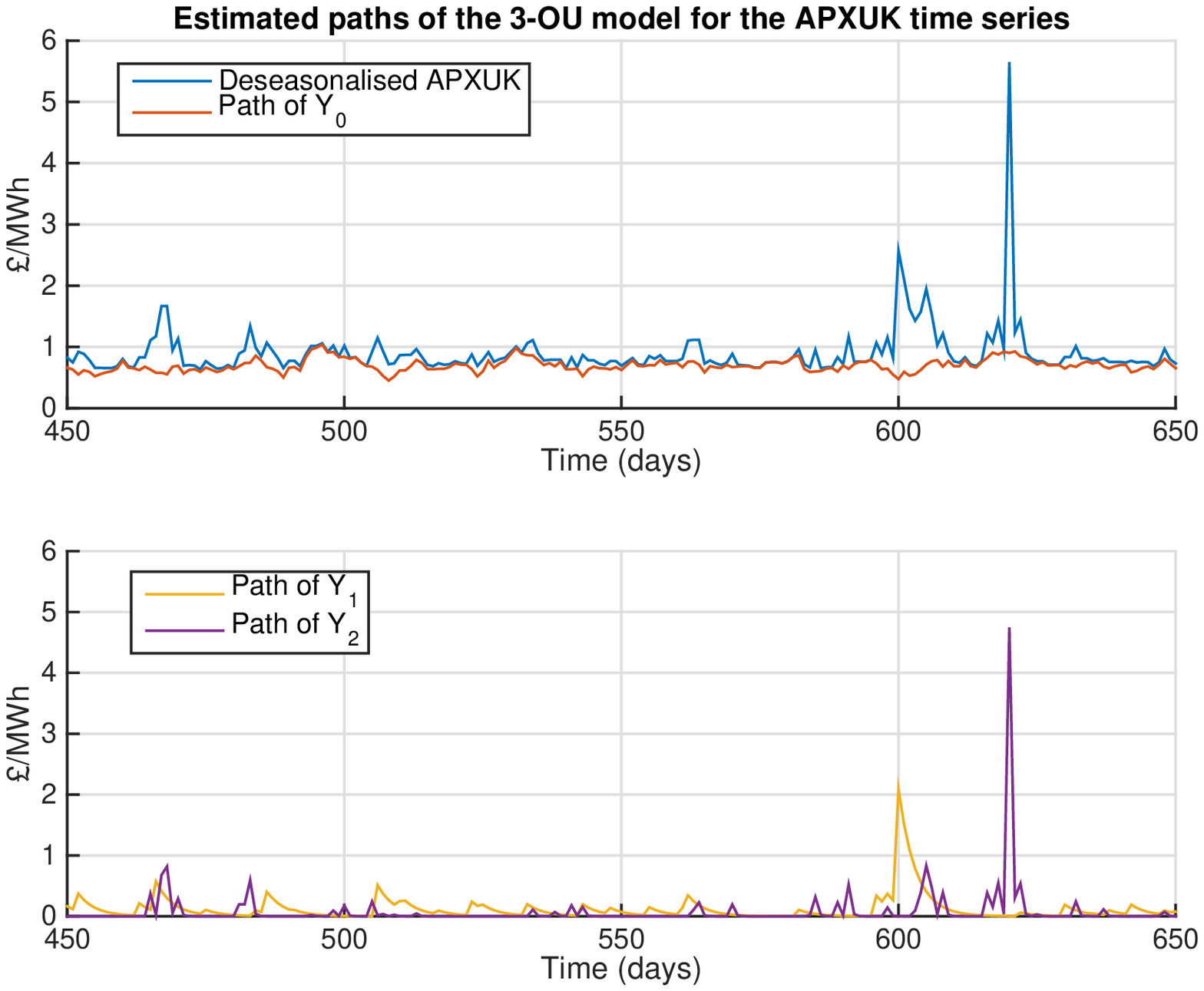}
		\par\end{centering}
	
	\caption{\label{fig:2c-APXUK-path-y0y1}Samples from the final state of the Markov chain for the jump processes, plus a section of the deseasonalised 2001--2006 APXUK time series superimposed on the implied diffusion process $Y_0$. Top: 2-OU model, bottom: 3-OU model.}
\end{figure}

In order to illustrate the role of the latent variables introduced in the augmented state space of our Markov chain, and to show how these may vary between the 2-OU and 3-OU models, Figure \ref{fig:2c-APXUK-path-y0y1} gives a representation of one posterior sample for each model via their respective latent processes $Y_i$. For clarity of the plot a restricted period of $200$ days is shown, giving the paths of the jump processes, plus the deseasonalised APXUK price superimposed on the implied diffusion $Y_0 = X - \sum_{i=1}^n Y_i$.
(For both models these samples are in fact the last state of the simulated Markov chain.)

From Table $\ref{tab:2OU-APXUK-prior-and-estimates}$ the jumps of $Y_1$ are relatively large (their distributional mean size $\beta$ has posterior expected value $0.72$) and the decay rate $\lambda_1$ has posterior mean $0.54$. The 3-OU model identifies both slowly decaying small jumps and rapidly decaying large jumps. Inspecting the plots in Figure \ref{fig:2c-APXUK-path-y0y1} for the 2-OU model around day $600$, it is therefore apparent that in this illustrative example runs of consecutive quickly decaying jumps combine to produce an apparently larger and more slowly decaying disturbance, while some single jumps such as that around day $630$ yield quickly decaying large spikes. In contrast the runs of overlapping spikes are much reduced in the plots for the 3-OU model.

\subsubsection{Diagnostics}

The posterior predictive $p$-values for the diffusion process $Y_0$ are $0.0617$ and $0.335$ for the 2-OU and 3-OU model respectively. Taking a posterior predictive $p$-value in excess of $0.1$ to be acceptable, two jump components are therefore required in order to give the diffusion process an acceptable fit on the basis of this diagnostic.

We also test the modelling assumption of Poisson jump arrivals with a constant intensity using the diagnostic described in Section \ref{sec:Diagnostics-theory}. 
The constant jump intensity model is acceptable for the APXUK data with a posterior predictive $p$-value for the distribution of spike inter-arrival times of approximately $0.4$, see Table \ref{tab:Bayesian-pvalues-times-sizes-APXUK-EEX}. We note finally that the exponential model for jump sizes is acceptable in all cases (with the posterior predictive $p$-value exceeding $0.3$). 

\begin{table}[htb]
	\begin{centering}
		\begin{tabular}{ccccccc}
						& \multicolumn{3}{c}{APXUK (2001-6)} & \multicolumn{3}{c}{EEX (2000-6)}\tabularnewline
			\cline{2-7} 
		Jump times of	& 2-OU & 3-OU &  & 2-OU & 3-OU & 3-OU-$I_1$ \tabularnewline
\hline
 $\Phi_{1}$ \qquad & 0.0525 & 0.4003 &  & 0.0099 & 0.0185 & 0.1643\tabularnewline
			 $\Phi_{2}$ \qquad & - - & 0.3089 &  & - - & 0.4681 & 0.4738\tabularnewline
			\hline 
		\end{tabular}
		\par\end{centering}
	
	\caption{\label{tab:Bayesian-pvalues-times-sizes-APXUK-EEX}
		Posterior predictive $p$-values for the model of jump times for processes $\Phi_{i}$.}

\end{table}

\subsection{2000--2006 EEX data}
We also calibrate 2-OU and 3-OU models to the 2000--2006 EEX data. Since exploratory analysis of the EEX dataset suggests the presence of frequent negative price spikes, our 3-OU model for the EEX series will differ from that for the APXUK dataset by specifying a negative sign for the second jump component $Y_2$.
Indeed, calibration of the 3-OU model with two positive jump components yields a posterior predictive $p$-value for the increments of the process $Y_0$ less than $0.005$ and the posterior distributions for the parameters of the two positive jump components are not well separated (data not shown), indicating that the {\em positive} price spikes in the EEX market tend to be driven by one jump component.
In the 3-OU model of \eqref{eqn:superposition_model} we therefore set $w_{0}=w_{1}=1$ and $w_{2}=-1$.

Further, taking into account the experience of past studies we consider both constant and periodic jump rates for the positive jump process $L_1(t)$, taking the periodic intensity function $I_1(\vartheta_1,t)$ given in \eqref{eq:periodic-intensity-rate} with $k=130$ days (which corresponds to a period of one half-year). We refer to the latter model as the 3-OU-$I_1$ model.
We take the same priors as in the APXUK studies above, now removing the restriction on the decay rates so that the prior for $\rho_2$ (or equivalently for $\lambda_2$) is independent and distributed as that for $\rho_1$ (or $\lambda_1$), since with jumps of opposite direction there should be no issue of identifiability.  Further, for the 3-OU-$I_1$ model the priors for $\eta_1$ and $\delta_1$ are both Ga$(1,10)$, while for  $\theta_1$ a U$(65,195)$ prior is taken. 

\subsubsection{Number of jump components}

\begin{sidewaystable}
	\begin{onehalfspace}
		\begin{centering}
			\begin{tabular}{cccccccccccc}
				\hline 
				Parameter & \multicolumn{3}{c}{Prior properties} & \multicolumn{2}{c}{2-OU} & \multicolumn{2}{c}{2-OU-$I_{1}$} & \multicolumn{2}{c}{3-OU} & \multicolumn{2}{c}{3-OU-$I_{1}$}\tabularnewline
				\hline 
				& Prior & Mean & SD & Mean & SD & Mean & SD & Mean & SD & Mean & SD\tabularnewline
				\hline 
				$\mu$ & $\mbox{N}(1,20^{2})$ & 1 & 20 & 0.9978 & 0.0182 & 0.9954 & (0.0184) & 1.0171 & 0.0230 & 1.0146 & 0.0226\tabularnewline
				$\sigma^{2}$ & $\mbox{IG}(1.5,0.005)$ & 0.01 & - - & 0.0271 & 0.0014 & 0.0269 & (0.0014) & 0.0122 & 0.0013 & 0.0119 & 0.0012\tabularnewline
				$e^{-1/\lambda_{0}}$ & $\mbox{U}(0,1)$ & 0.5 & 0.2887 & 0.7978 & 0.0166 & 0.7980 & (0.0162) & 0.8821 & 0.0151 & 0.8835 & 0.0150\tabularnewline
				$(\lambda_{0})$ & ($\mbox{IG}(1,1)$) & (- -) & (- -) & (4.4612) & (0.4203) & (4.4635) & (0.4071) & (8.1130) & (1.1438) & (8.2193) & (1.1698)\tabularnewline
				$e{}^{-1/\lambda_{1}}$ & U$(0,1)$ & 0.5 & 0.2887 & 0.1057 & 0.0267 & 0.1142 & (0.0246) & 0.1915 & 0.0313 & 0.1809 & 0.0344\tabularnewline
				$(\lambda_{1})$ & ($\mbox{IG}(1,1)$) & (- -) & (- -) & (0.4441) & (0.0512) & (0.4603) & (0.0465) & (0.6060) & (0.0601) & (0.5859) & (0.0657)\tabularnewline
				$e^{-1/\lambda_{2}}$ & U$(0,1)$ & 0.5 & 0.2887 & - - & - - & - - & - - & 0.2295 & 0.0348 & 0.2230 & 0.0384\tabularnewline
				$(\lambda_{2})$ & ($\mbox{IG}(1,1)$) & (- -) & (- -) & - - & - - & - - & - - & (0.6814) & (0.0709) & (0.6687) & (0.0775)\tabularnewline
				$\eta_{1}$ & $\mbox{Ga}(1,10)$ & 0.1 & 0.1 & 0.1049 & 0.0185 & - - & - - & 0.1274 & 0.0192 & - - & - -\tabularnewline
				$\eta_{1}^{*}$ & $\mbox{Ga}(1,10)$ & 0.1 & 0.1 & - - & - - & 0.2881 & (0.0619) & - - & - - & 0.2515 & 0.0428\tabularnewline
				$\eta_{2}$ & $\mbox{Ga}(1,10)$ & 0.1 & 0.1 & - - & - - & - - & - - & 0.1438 & 0.0335 & 0.1422 & 0.0320\tabularnewline
				$\beta_{1}$ & $\mbox{IG}(1,1)$ & - - & - - & 1.0971 & 0.1435 & 1.0737 & (0.1401) & 0.9045 & 0.1088 & 0.8998 & 0.1052\tabularnewline
				$\beta_{2}$ & $\mbox{IG}(1,1)$ & - - & - - & - - & - - & - -  & - - & 0.4176 & 0.0734 & 0.4308 & 0.0793\tabularnewline
				$\theta_{1}$ & U$(65,195)$ & 130 & 37.5278 & - - & - - & 135.0048 & (2.8474) & - - & - - & 141.3725 & 3.5071\tabularnewline
				$\delta_{1}$ & $\mbox{Ga}(1,10)$ & 0.1 & 0.1 & - - & - - & 0.6271 & (0.1281) & - - & - - & 0.3408 & 0.0884\tabularnewline
				\hline 
			\end{tabular}
			\par\end{centering}
	\end{onehalfspace}	
	\caption{\label{tab:prior-3OU-EEX-5Lup-1}Prior distributions and posterior properties
			when fitting the 2- and 3-OU models to the 2000--2006 EEX data. ${}^*$ In the 2-OU-$I_1$ and 3-OU-$I_1$ models the parameter $\eta_1$ indicates the maximum jump rate of the periodic intensity function $I_1$. The
			third OU component $Y_{2}$ of the models 3-OU and 3-OU-$I_1$ is negative. The indirect parameters $\lambda_i$ are treated as described in the caption to Table \ref{tab:2OU-APXUK-prior-and-estimates}.	
		}
\end{sidewaystable}

Table \ref{tab:prior-3OU-EEX-5Lup-1} presents summary statistics for the posterior distributions of the parameters for both the 2-OU model and the 3-OU models applied to the 2000--2006 EEX data. The posterior diagnostic for the diffusion component $Y_0$ is acceptable for both three-component models while, as in the APXUK case, the two-component model does not appear to be satisfactory: the posterior $p$-value for $Y_0$ is equal to $0.0021$ for the 2-OU model and equal to $0.192$ and $0.26$ for the 3-OU and 3-OU-$I_1$ models respectively. This is explained by the fact that negative price jumps are not accounted for with the 2-OU model, frequently resulting in large residuals for $Y_0$. 

There is agreement across the first two moments of the posterior distributions for all parameters common to the 3-OU and 3-OU-$I_1$ models.
In contrast with the 2001--2006 APXUK dataset, however, the results for EEX support the presence of seasonality in the occurrence of price spikes, see Table \ref{tab:Bayesian-pvalues-times-sizes-APXUK-EEX}. The constant jump intensity model appears to be unsatisfactory for the first jump component $\Phi_1$, with a corresponding $p$-value of $0.0185$ in the 3-OU model, while the 3-OU-$I_1$ returns a $p$-value of $0.1643$.
Figure \ref{3c-EEX-average-number-pos-jumps} displays the number of positive jumps on the EEX market by month, averaged over our posterior samples of the process $\Phi_1$ in the 3-OU-$I_1$ model.

\subsection{2011 - 2015 data}\label{subsec:2011-15}

\begin{table}[tb]
	\begin{centering}
		\begin{tabular}{cccccc}
			\hline 
			& APXUK (2011-15)&  & \multicolumn{3}{c}{EEX (2011-15)}\tabularnewline
			\cline{2-2} \cline{4-6} 
			& 2-OU &  & 2-OU & 2-OU-$I_{1}$ & 2-OU$^{-}$\tabularnewline
			\hline 
			$Y_{0}$ & 0.2167 &  & 0.0001 & 0.0012 & 0.1502\tabularnewline
			Jump times of $\Phi_{1}$ & 0.3521 &  & 0.4219 & 0.3207 & 0.4452\tabularnewline
			Jump sizes of $\Phi_{1}$ & 0.4574 &  & 0.5121 & 0.4554 & 0.4998\tabularnewline
			\hline 
		\end{tabular}
		\par\end{centering}
	
	\caption{\label{tab:Bayesian-pvalues-times-sizes-APXUK-EEX-1-1-1}Posterior predictive	$p$-values for a range of models for the APXUK and EEX indices over the sample period January 24, 2011 to February 2, 2015.}
\end{table}

Motivated by visual inspection of the price data in Figure \ref{fig:deseasonalised-data}, as discussed in Section \ref{sec:motivation}, we wish to examine whether the statistical structure of the price data differs in periods before and after the global financial crises of 2007-8 and 2009. The models given by \eqref{eqn:superposition_model} were therefore calibrated to the APXUK and EEX indices over the sample period ranging from January 24, 2011 to February 16, 2015, and the simplest acceptable models were identified on the basis of  posterior predictive $p$-values (again taking 0.1 as the minimum acceptable level). It may be seen from Table \ref{tab:Bayesian-pvalues-times-sizes-APXUK-EEX-1-1-1} that the 2011-2015 APXUK data supports the 2-OU model with one positive jump component. In order to discuss the statistics of the jump processes we will refer to the posterior mean values presented in Table \ref{tab:prior-3OU-APXUK} as `Old' and in Table \ref{tab:MCMC-realdata-2010-2015} as `New'. Although both the `New' values lie between the corresponding `Old' values,
$\beta_1^{\text {Old}} < \beta_1^{\text{New}}<\beta_2^{\text{Old}}$ and $\lambda_1^{\text {Old}} < \lambda_1^{\text{New}}<\lambda_2^{\text{Old}}$, the new jump process cannot be interpreted as simply a statistical mixture of the two old jump processes since its intensity is lower than both of the old jump intensities.\footnote{Also, the sum of two jump OU processes with different mean reversion rates is statistically significantly different from one jump OU process and cannot, therefore, be successfully approximated by the latter.} Indeed, on average the total number of jumps (of any size) per unit time is less than a third for the 2011-2015 data compared to 2001--2006.

From Table \ref{tab:Bayesian-pvalues-times-sizes-APXUK-EEX-1-1-1}, the 2011-2015 EEX data in fact supports the 2-OU$^-$ model which has a single, negative jump component (motivating the superscript minus in the notation). In this model the small number of larger upward price movements above the mean level in Figure \ref{fig:deseasonalised-data} must therefore be accounted for by correspondingly large residuals in the diffusion component $Y_0$, and this explains the relatively low predictive $p$-value (0.1502) for $Y_0$. With regard to the statistics of the negative jump component, the values $\lambda_1, \eta_1, \beta_1$ in Table \ref{tab:MCMC-realdata-2010-2015} should be compared to the values of $\lambda_2, \eta_2, \beta_2$ in Table \ref{tab:prior-3OU-EEX-5Lup-1}. Since negative prices were introduced in this market on September 1, 2008 \citep{genoese2010occurrence}, in general larger negative jumps were possible in the 2011-2015 data. Indeed the most significant downward jump in 2011-2015 was to a large negative price, and our finding $0.593=\beta_1^\text{New}>\beta_2^\text{Old}=0.4308$ is consistent with this change to the EEX market structure.
For the diffusion component, the coefficient $\lambda_0$ decreases from 11.9 and 8.22 for the APXUK and EEX markets respectively to approximately 3.6, a value which happens to be consistent across both markets. 

\begin{table}[tb]
	\begin{onehalfspace}
		\begin{centering}
			\begin{tabular}{ccccccc}
				\cline{3-7} 
				&  & \multicolumn{2}{c}{APXUK (2011-15)} &  & \multicolumn{2}{c}{EEX (2011-15)}\tabularnewline
				\cline{3-7} 
				&  & \multicolumn{2}{c}{2-OU model} &  & \multicolumn{2}{c}{2-OU$^{-}$ model}\tabularnewline
				\cline{3-7} 
				Parameter &  & Mean & SD &  & Mean & SD\tabularnewline
				\hline 
				$\mu$ &  & 0.9865 & 0.0095 &  & 1.0480 & 0.0149\tabularnewline
				$\sigma^{2}$ &  & 0.0071 & 0.0006 &  & 0.0170 & 0.0016\tabularnewline
				$e^{-1/\lambda_{0}}$ &  & 0.7537 & 0.0232 &  & 0.7590 & 0.0251\tabularnewline
				$(\lambda_{0})$ &  & (3.5727) & (0.3976) &  & (3.6736) & (0.4526)\tabularnewline
				$e{}^{-1/\lambda_{1}}$ &  & 0.2104 & 0.0394 &  & 0.1941 & 0.0381\tabularnewline
				$(\lambda_{1})$ &  & (0.6435) & (0.0781) &  & (0.6115) & (0.0741)\tabularnewline
				$\eta_{1}$ &  & 0.1172 & 0.0324 &  & 0.1105 & 0.0310\tabularnewline
				$\beta_{1}$ &  & 0.3981 & 0.0921 &  & 0.5930 & 0.1325\tabularnewline
				\hline 
			\end{tabular}
			\par\end{centering}
	\end{onehalfspace}
	
	\caption{\label{tab:MCMC-realdata-2010-2015}Posterior properties obtained when calibrating the 2-OU (one positive jump component) and 2-OU$^{-}$ (one negative jump component) models to the APXUK and EEX datasets respectively. The sample period is January 24, 2011 to February 16, 2015.}
\end{table}

\section{Discussion}\label{sec:discussion}

\subsection{Scope of contribution}

We have shown that multiple components can be required to obtain statistically adequate mean-reverting models of electricity spot prices and, further, that the required combination of components can change over time. To clarify the value of this contribution we note that spot price models
have two principal areas of application in the literature. Firstly, price forecasting is concerned with the prediction of prices over future time points or periods given the current and past values of relevant variables (see for example \cite{weron_modeling_2007}). As such it is not concerned with the detailed statistical properties of spot price \emph{trajectories} such as the long-run statistical patterns in price spikes, which are the object of our work. Instead studies (such as ours) of spot price dynamics are suitable both for derivative pricing \citep{hull_options_2009} and operational analyses in the real options framework (see for example \cite{kitapbayev2015stochastic, moriarty2017real}). In derivative pricing a main goal is to combine the latter models with observed derivative prices to construct so-called {\em risk-neutral} or {\em martingale} probability measures. Thus while model parameters are inferred from derivative prices it is important to identify the right class of price models, and our work provides an approach to this question via posterior predictive checking. In contrast, in real options analyses the fact that real projects are not traded means that the {\em physical} or {\em historic} probability measure is often the one used. In this context our work provides an approach both to model specification and to the calibration of model parameters to historic data.

The methodological advantages of our approach to calibration, which aims to make minimal assumptions about the spike processes, are confirmed by the 2000--2006 APXUK analysis. In mean-reverting models jumps do not immediately vanish but instead decay over time. This means that jump components, particularly those having the same direction, can interact when superposed. Inference on the individual spike components is then more challenging and simple signal processing approaches (c.f. \cite{meyer-brandis_multi-factor_2008}), such as the use of thresholds to identify jumps, are rendered unsuitable. Nevertheless we have shown that a statistically adequate model can be extracted. The ability of our MCMC procedure to distinguish spikes in the same direction is confirmed in the appendix using simulated price data. There, Figure A.2 plots the simulated jumps (in red) and a visual representation of the posterior distribution of the latent jump processes (blue, see Section A.3 for details) so that the agreement can be assessed.

\subsection{Fundamental drivers}

In this section we attempt to relate 
our empirical results to their underlying physical and economic drivers. Negative price spikes are associated with the priority given to wind energy in the spot market \citep{benth_stochastic_2013}. A glut in wind power production can lead to a corresponding  decrease in demand for other sources of generation. It can be impossible for conventional generators to reduce production sufficiently so they may temporarily 
accept low (or even negative) prices. In support of this analysis, we observed that the inclusion of a negative jump component was necessary for adequate modelling of the EEX data in both periods. Further the negative jumps had higher mean size in the later period, a change consistent with the increasing penetration of renewable generation.


While preprocessing the data we removed a deterministic seasonal component from the spot prices. However in some electricity markets (particularly in the US and Europe) seasonality has also been observed in the frequency of price spikes \citep{geman_understanding_2006, benth_critical_2012}. A priori this may be explained by greater levels of stress in the power system during the extremes of seasonal variation in weather due, for example, to heating load during cold snaps. The presence of jump seasonality was suggested in the 2000--2006 EEX data as illustrated in Figure \ref{3c-EEX-average-number-pos-jumps}, which displays the number of positive jumps on the EEX market by month, averaged over samples from our MCMC procedure. 
(It should be noted that Figure \ref{3c-EEX-average-number-pos-jumps} is indicative and does not represent direct statistical estimates of jump frequency in spot prices.) Indeed for the latter series it was necessary to incorporate seasonality in the arrival rate of the positive jump component in order to obtain a statistically adequate model. In contrast seasonal jump components were not statistically necessary for the APXUK data in either period, which may be related to the less severe extremes of UK winter weather.
 
As presented in Section \ref{subsec:summ}, there is statistical evidence for a reduction across both markets in both the number of positive jump components and the frequency of positive price spikes after the global financial crises of 2007-8 and 2009. It is true that in both the UK and Germany, electricity consumption generally increased in the period 2000--2006 and was generally level or decreased during 2011--2015.\footnote{\url{http://data.worldbank.org}}
It follows that the power systems under study faced less stress from constraints in either production or transmission capacity during the latter period, and this is consistent with the observed reduction in positive price spikes. It must be noted however that the outlook for the future is somewhat different, with changes on the supply side including the decommissioning of ageing and carbon-intensive conventional generation and increased reliance on intermittent generation (see, for example, \cite{grid_etys}), suggesting that positive spikes may return to the electricity spot market.

\section{Conclusions}
\label{sec:conclusion}

By modelling mean-reverting deseasonalised electricity spot prices as the sum of a diffusion process and multiple signed jump processes of deterministic intensity, and applying a Bayesian calibration procedure and posterior diagnostics, we have identified a class of  multi-factor  models suitable for modelling empirical prices across two different markets and two different time periods. In contrast with several recent studies using stochastic volatility models we have employed multiple signed jump components, albeit with simpler deterministic volatilities (either constant, or deterministic and periodic). This approach allows straightforward comparison of the statistical structure of prices across different markets and time periods: each model has a number of signed jump components with distinguished jump intensities, decay rates and size distributions. In both the APXUK and EEX markets it was found that the statistical structure of the price series differs before and after the period 2007-2010 and, in particular, that the number of positive jump components decreased (from 2 to 1 and 1 to 0 respectively), with the mean reversion speed of the diffusive price component increasing 
in both markets. Seasonality in the jump intensity was found to be necessary only in the earlier (2000--2006) EEX data and only for its positive jump component.

\section*{Acknowledgements}
JM was supported by grants EP/K00557X/1 and EP/K00557X/2 and JG was supported by grant EP/I031650/1 from the UK Engineering and Physical Sciences Research Council. JP was supported in part by MNiSzW grant UMO-2012/07/B/ST1/03298.




\bibliographystyle{apalike}

\bibliography{mylibrary}

\appendix
\setcounter{figure}{0}
\setcounter{table}{0}
\renewcommand\appendixname{}
\label{sec:Appendix}

\section{Notes on implementation}

In the first three sections of this Appendix we explore in detail selected aspects of our inference procedure by the use of simulated data. In Section \ref{sec:simul} we examine the balance between jumps and diffusion in the model, since very frequent jumps may potentially be accumulated into the diffusion component during inference. Then in Section \ref{sec:multiple} we explain how a particular implementation issue in the 3-OU model was addressed, where the presence of two positive jump components resulted in  slow mixing. In Section \ref{sec:repn} we illustrate the algorithm's performance in estimating the high-dimensional state of the latent jump process. Further analysis of our MCMC algorithm can be found in \citet[Chapter 5]{gonzalez_modelling_2015}, where  extensive testing for the 2-OU and 3-OU models on simulated data is carried out. 

Section \ref{subsec:moments_rho_2} explores the prior distribution of $\rho_2$ when there are two jump components of the same sign and an ordering between $\lambda_1$ and $\lambda_2$ (and consequently between $\rho_1$ and $\rho_2$) is imposed through the joint prior. The final two sections provide further details of the case studies of Section \ref{sec:real}, namely deseasonalisation of the raw data (Section \ref{sec:fit-of-seasonal-trend-function}) and the sensitivity of the results to the choices of prior distributions  (Section \ref{sec:psa}).

\subsection{Dependence of posterior distributions on $\eta$}
\label{sec:simul}

To explore the influence of actual jump rates on the output posterior distributions we simulate daily data for 1000 days from the model in
equation (\ref{eqn:superposition_model}) with $n=1$ and a range of constant jump intensities $\eta$, corresponding to averages from 13 to 78 jumps per year, with all other parameters fixed as in Table \ref{tab:simulation-study-true-values-priors}. This range of jump intensities has been reported in the literature for energy spot price models (cf. \citet{seifert_modelling_2007,meyer-brandis_multi-factor_2008,benth_stochastic_2008,benth_critical_2012}). Taking the prior distributions listed in Table \ref{tab:simulation-study-true-values-priors} we then apply our MCMC algorithm. Table \ref{tab:simulation-study} summarises the results. It may be seen that the accurate separation between the jump process and the diffusion, as measured by the posterior moments of the diffusion coefficient $\sigma$, is maintained even with a large number of jumps. Furthermore there is generally a negligible influence of the jump intensity on the posterior distributions of other parameters. The main exception is the posterior distribution of the jump size parameter $\beta$, which becomes more concentrated with its mean closer to the simulation value with increasing values of $\eta$, the result of more informative data (more jumps) being available for estimation.

\begin{table}[p]
\begin{onehalfspace}
\begin{centering}
	\begin{tabular}{ccccc}
		\hline 
		&  & \multicolumn{3}{c}{Prior properties}\tabularnewline
		\hline 
		Parameter & Simulation value & Prior & Mean & SD\tabularnewline
		\hline 
		$\mu$ & 1 & $\mbox{N}(1,20^{2})$ & 1 & 20\tabularnewline
		$\sigma^{2}$ & 0.01 & $\mbox{IG}(1.5,0.005)$ & 0.01 & - -\tabularnewline
		$e^{-1/\lambda_0}$ & $e^{-1/8}\approx0.8825$ & $\mbox{U}(0,1)$ & 0.5 & $0.2887$\tabularnewline
		$(\lambda_{0})$ & (8) & - - & - - & - -\tabularnewline
		$e^{-1/\lambda_1}$ & $e^{-1/2}\approx0.6065$ & $\mbox{U}(0,1)$ & 0.5 & $0.2887$\tabularnewline
		$(\lambda_1$) & (2) & - - & - -  & - - \tabularnewline
		$\eta$ & $\{0.05,0.1,0.2,0.3\}$ & $\mbox{Ga}(1,\eta_{true}^{-1})$ & $\eta_{true}$ & $\eta_{true}$\tabularnewline
		$\beta$ & 0.7 & $\mbox{IG}(1,1)$ & - - & - -\tabularnewline
		\hline 
	\end{tabular}
	\par\end{centering}
\end{onehalfspace}

\caption{\label{tab:simulation-study-true-values-priors}Parameter values used to generate simulated data and the prior distributions used in the calibration exercise. Where specified in the prior distribution, $\eta_{true}$ takes the value of the jump intensity $\eta$ used in the simulation.}
\end{table}

\begin{table}[p]
	\begin{onehalfspace}
		\begin{centering}
			\begin{tabular}{ccccccccc}
				\hline 
				\multicolumn{2}{c}{True value} & \multicolumn{1}{c}{$\eta=0.05$} &  & \multicolumn{1}{c}{$\eta=0.1$} & \multicolumn{1}{c}{} & \multicolumn{1}{c}{$\eta=0.2$} &  & \multicolumn{1}{c}{$\eta=0.3$}\tabularnewline
				\hline 
				$\mu$ & 1 & 0.9999 &  & 0.9959 &  & 0.9954 &  & 0.9988\tabularnewline
				&  & {\footnotesize{}(0.9527, 1.0471)} &  & {\footnotesize{}(0.9535, 1.0383)} &  & {\footnotesize{}(0.9404, 1.0504)} &  & {\footnotesize{}(0.9452, 1.0524)}\tabularnewline
				$\sigma^{2}$ & 0.01 & 0.0101 &  & 0.0101 &  & 0.0100 &  & 0.0101\tabularnewline
				&  & {\footnotesize{}(0.009, 0.0112)} &  & {\footnotesize{}(0.009, 0.0111)} &  & {\footnotesize{}(0.0087, 0.0113)} &  & {\footnotesize{}(0.0088, 0.0115)}\tabularnewline
				$e^{-1/\lambda_{0}}$ & 0.8825 & 0.8806 &  & 0.8820 &  & 0.8767 &  & 0.8819\tabularnewline
				&  & {\footnotesize{}(0.8457, 0.9155)} &  & {\footnotesize{}(0.854, 0.91)} &  & {\footnotesize{}(0.843, 0.9104)} &  & {\footnotesize{}(0.8506, 0.9132)}\tabularnewline
				$(\lambda_{0})$ & (8) & 8.0463 &  & 8.0892 &  & 7.7580 &  & 8.1052\tabularnewline
				&  & {\footnotesize{}(5.5261, 10.5664)} &  & {\footnotesize{}(6.0167, 10.1618)} &  & {\footnotesize{}(5.4788, 10.0372)} &  & {\footnotesize{}(5.8665, 10.3439)}\tabularnewline
				$e^{-1/\lambda_1}$ & 0.6065 & 0.6082 &  & 0.6074 &  & 0.6084 &  & 0.6077\tabularnewline
				&  & {\footnotesize{}(0.5755, 0.6409)} &  & {\footnotesize{}(0.5808, 0.6339)} &  & {\footnotesize{}(0.5874, 0.6293)} &  & {\footnotesize{}(0.5908, 0.6246)}\tabularnewline
				$(\lambda_1)$ & (2) & 2.0157 &  & 2.0086 &  & 2.0140 &  & 2.0087\tabularnewline
				&  & {\footnotesize{}(1.7945, 2.237)} &  & {\footnotesize{}(1.8267, 2.1905)} &  & {\footnotesize{}(1.8736, 2.1543)} &  & {\footnotesize{}(1.8963, 2.1212)}\tabularnewline
				$\eta$ & - - & 0.0474 &  & 0.0982 &  & 0.1954 &  & 0.2993\tabularnewline
				&  & {\footnotesize{}(0.0279, 0.067)} &  & {\footnotesize{}(0.0698, 0.1265)} &  & {\footnotesize{}(0.1522, 0.2385)} &  & {\footnotesize{}(0.2547, 0.344)}\tabularnewline
				$\beta$ & 0.7 & 0.7915 &  & 0.7455 &  & 0.7302 &  & 0.7151\tabularnewline
				&  & {\footnotesize{}(0.5398, 1.0433)} &  & {\footnotesize{}(0.5784, 0.9126)} &  & {\footnotesize{}(0.5937, 0.8667)} &  & {\footnotesize{}(0.5891, 0.8412)}\tabularnewline
				\hline 
			\end{tabular}
			\par\end{centering}
	\end{onehalfspace}
	
	\caption{\label{tab:simulation-study}Average and spread of posterior means across 60
		runs of the MCMC algorithm as the simulation value of $\eta$ varies. The intervals shown represent this average plus and minus 1.96 standard deviations of the posterior mean values.
		The indirect parameters $\lambda_i$ are treated
		as described in the caption to Table \ref{tab:2OU-APXUK-prior-and-estimates}.}
\end{table}

\subsection{Multiple updates of the latent process}
\label{sec:multiple}
\begin{figure}[p]
\begin{centering}
\includegraphics[scale = 0.6]{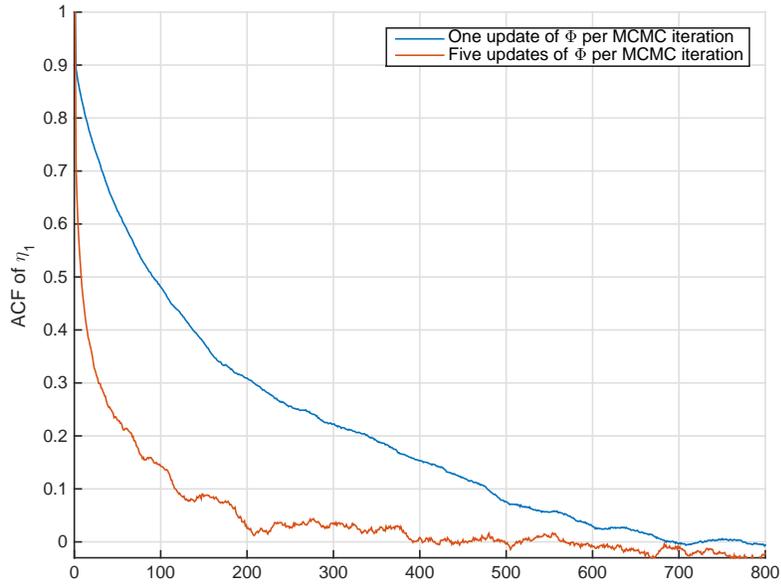}
\par\end{centering}

\caption{\label{fig:3c-sim-acf-eta-1vs5Lupdates}The autocorrelation function of $\eta_{1}$ when
fitting the 3-OU model with two positive jump components to simulated data, using one (blue line) and
five (orange line) updates of the latent process $\Phi$ per update
of the remaining parameters.}
\end{figure}

The dimensionality of the latent process is much higher than that of the other variables (model parameters) targeted by the Markov chain described in Subsection \ref{sec:MCMC_implementation} (it is potentially infinite). A single update of the jump process $\Phi_i$ can affect as little as one jump and therefore can have a much smaller effect on the process than a single update in any other step of the Gibbs sampler. This results in slow mixing of the chain, which becomes most pronounced in the 3-OU model. In this case the algorithm is therefore modified so that the updates to both $\Phi_{1}$ and $\Phi_{2}$ described in Section \ref{sec:phiupdate} are applied five times per single MCMC iteration. This modification results in significant improvements to observed mixing for all model parameters. As an illustration, Figure \ref{fig:3c-sim-acf-eta-1vs5Lupdates} provides the autocorrelation function (ACF) of
$\eta_{1}$ when using the original and the modified schemes for updating $\Phi$.

\subsection{Jump process posteriors}
\label{sec:repn}

\begin{figure}[p]
\includegraphics[width=\textwidth]{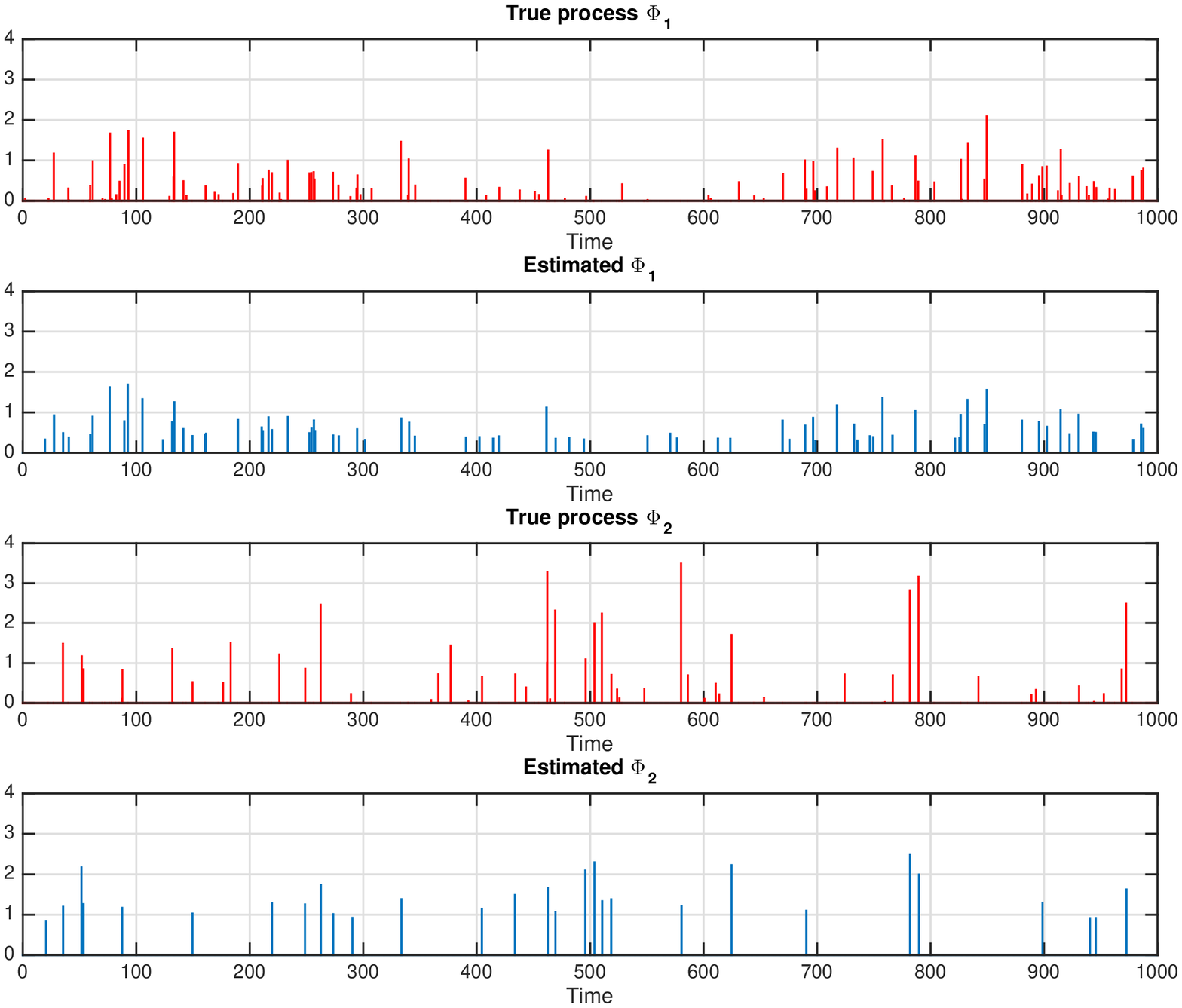}
\caption{\label{fig:3c-true-and-estimated-L}The simulated jump processes $\Phi_{1}$ 
and $\Phi_{2}$ for the 3-OU algorithm (red) and a representation of the last 5000 states of the jump processes in the Markov chain (blue, for details see Appendix \ref{sec:repn}).}
\end{figure}

In order to illustrate the posterior distributions obtained for the latent jump processes $\Phi_1$ and $\Phi_2$, Figure \ref{fig:3c-true-and-estimated-L} presents results from a simulation study with the 3-OU model. Data was generated from model \eqref{eqn:superposition_model} with $n=2$ with the following parameter values:
\[ 
(\mu,\lambda_{0},\lambda_{1},\lambda_2,\sigma,\beta_1,\beta_2,\eta_1,\eta_2)=(1,8,3,0.5,0.15,0.5,1,0.1,0.05).
\]

Our 3-OU MCMC algorithm was then applied taking the priors in Table \ref{tab:prior-3OU-APXUK}. Figure \ref{fig:3c-true-and-estimated-L} provides a representation of the last 5000 states of the jump processes in the Markov chain, as follows. For each day $j$ having one or more jumps in at least 3000 of these states, the observed jump sizes were averaged;  in states where there was more than one jump on that day, the sum of these jump sizes was taken. This average observed jump size $\bar{\xi}_j$ was then plotted against day $j$.

\subsection{Prior moments of $\rho_2$}\label{subsec:moments_rho_2}
When both jump components have the same sign we impose the condition $\lambda_1 > \lambda_2$ via specification of the prior: $\rho_2 \sim \rho_1 U(0,1)$. The resulting distribution of $\rho_2$ is non-standard with the cumulative distribution function
\[
\mathbb{P}(\rho_2 \le x) = \int_0^1 \Big(\frac{x}{\rho_1} \wedge 1\Big) d\rho_1 = x (1 - \log(x)), \quad x \in (0,1).
\]
Using integration by parts we compute the first moment $\mathbb{E} (\rho_2) = 1/4$ and the second moment $\mathbb{E} (\rho^2_2) = 1/9$.

\subsection{Deseasonalisation}\label{sec:fit-of-seasonal-trend-function}

For completeness Table \ref{tab:estimated-seasonal-trend} presents coefficients from fitting the seasonal trend in \eqref{eq:seasonal-trend}. In the case of 2011-2015 EEX data, which contains three instances of negative prices, these were replaced by averages of the neighbouring price values for the purposes of deseasonalisation only.

\begin{table}[p]
	\begin{onehalfspace}
	\begin{centering}
		\begin{tabular}{cccccccc}
			\hline 
			Dataset &  & $a_{1}$ & $a_{2}$ & $a_{3}$ & $a_{4}$ & $a_{5}$ & $a_{6}$\tabularnewline
			\hline 
			APXUK (2001-6) &  & 2.5770 & 0.0008 & -0.0817 & 0.0443 & -0.0097 & -0.0395\tabularnewline
			EEX (2000-6) &  & 2.9399 & 0.0006 & 0.0055 & -0.0803 & 0.0415 & -0.0140\tabularnewline
			APXUK (2011-15) &  & 3.9005 & -0.0001 & -0.0014 & 0.0342 & 0.0104 & -0.0368\tabularnewline
			EEX (2011-15) &  & 4.0399 & -0.0005 & -0.0585 & 0.0156 & 0.0298 & -0.0315\tabularnewline
			\hline 
		\end{tabular}
	\par\end{centering}
	\end{onehalfspace}
	
	\caption{\label{tab:estimated-seasonal-trend}Fitted seasonal trend coefficients in daily units.}

\end{table}

\subsection{Prior sensitivity analysis}
\label{sec:psa}
Our aim in the empirical studies of Section \ref{sec:real} was to let the data speak for itself. In order to explore the degree of sensitivity of the above results to the choice of prior distributions we apply the 3-OU-$I_1$ algorithm, which is the most complex of the above algorithms, to the 2000-2006 EEX  dataset. Among the large number of parameters we choose to vary the priors of $\sigma^2,\eta_{1}$ and $\eta_{2}$, since these parameters proved to exhibit relatively slow mixing, which might point to difficulties with estimation.
We make the following variations to the set of priors (replacing the corresponding Gibbs steps with Metropolis-Hastings steps as necessary):
\begin{itemize}
	\item Prior 1: as in Table \ref{tab:prior-3OU-EEX-5Lup-1}
	\item Prior 2: as in Table \ref{tab:prior-3OU-EEX-5Lup-1}, except that $\sigma^{2}\sim\mbox{U}(0,0.25^{2})$,
	\item Prior 3: as in Table \ref{tab:prior-3OU-EEX-5Lup-1}, except that $\pi(\eta_{i})\propto 1_{\{\eta_i>0\}}, i=1,2$ \emph{(improper uninformative priors)}.
\end{itemize}
Table \ref{tab:3c-EEX-sensitivity} presents the first two moments of the posterior distributions under these alternative sets of priors. The variations in these moments are insignificant, indicating that the data provides a clear indication of the parameter values.

\begin{table}[btp]
	\begin{onehalfspace}
	\begin{centering}
		\begin{tabular}{cccccccccc}
			\hline 
			& \multicolumn{9}{c}{Posterior properties}\tabularnewline
			\cline{3-10} 
			&  & \multicolumn{2}{c}{Prior 1} &  & \multicolumn{2}{c}{Prior 2} &  & \multicolumn{2}{c}{Prior 3}\tabularnewline
			\cline{3-10} 
			Parameter &  & Mean & SD &  & Mean & SD &  & Mean & SD\tabularnewline
			\hline 
			$\mu$ &  & 1.0146 & 0.0226 &  & 1.0144 & 0.0224 &  & 1.0149 & 0.0222\tabularnewline
			$\sigma^{2}$ &  & 0.0119 & 0.0012 &  & 0.0123 & 0.0013 &  & 0.0123 & 0.0014\tabularnewline
			$e^{-1/\lambda_{0}}$ &  & 0.8835 & 0.0150 &  & 0.8804 & 0.0151 &  & 0.8809 & 0.0158\tabularnewline
			$\lambda_{0}$ &  & 8.2193 & 1.1698 &  & 7.9874 & 1.1095 &  & 8.0406 & 1.1894\tabularnewline
			$e{}^{-1/\lambda_{1}}$ &  & 0.1809 & 0.0344 &  & 0.1905 & 0.0327 &  & 0.1826 & 0.0314\tabularnewline
			$\lambda_{1}$ &  & 0.5859 & 0.0657 &  & 0.6043 & 0.0628 &  & 0.5889 & 0.0601\tabularnewline
			$e^{-1/\lambda_{2}}$ &  & 0.2230 & 0.0384 &  & 0.2282 & 0.0374 &  & 0.2481 & 0.0446\tabularnewline
			$\lambda_{2}$ &  & 0.6687 & 0.0775 &  & 0.6791 & 0.0764 &  & 0.6730 & 0.0733\tabularnewline
			$\eta_{1}$ &  & 0.2515 & 0.0428 &  & 0.2492 & 0.0431 &  & 0.2253 & 0.0362\tabularnewline
			$\eta_{2}$ &  & 0.1422 & 0.0320 &  & 0.1381 & 0.0308 &  & 0.1329 & 0.0223\tabularnewline
			$\beta_{1}$ &  & 0.8998 & 0.1052 &  & 0.8928 & 0.1129 &  & 0.9105 & 0.1169\tabularnewline
			$\beta_{2}$ &  & 0.4308 & 0.0793 &  & 0.4272 & 0.0792 &  & 0.4329 & 0.0671\tabularnewline
			$\theta_{1}$ &  & 141.3725 & 3.5071 &  & 140.6855 & 3.5296 &  & 140.8370 & 3.5109\tabularnewline
			$\delta_{1}$ &  & 0.3408 & 0.0884 &  & 0.3531 & 0.0892 &  & 0.3546 & 0.0899\tabularnewline
			\hline 
		\end{tabular}
		\par\end{centering}
	\end{onehalfspace}
	
	\caption{\label{tab:3c-EEX-sensitivity}Prior sensitivity analysis for the 3-0U-$I_1$ model.}
\end{table}

\end{document}